\documentclass[footinbib,twocolumn,superscriptaddress,longbibliography]{revtex4-1}

\usepackage{graphicx}
\usepackage{latexsym}
\usepackage{graphics}
\usepackage{bm}
\usepackage[normalem]{ulem}
\usepackage{graphicx} 
\usepackage{color}
\usepackage{amsmath}
\usepackage{amssymb}
\usepackage{feynmp-auto}
\usepackage{slashed}
\usepackage{enumerate}
\usepackage{amsfonts, amssymb,amsmath,latexsym,graphics, graphicx,epsfig,multirow,comment,
,feyn,slashed,xcolor,afterpage, makecell} 
\usepackage{booktabs, blindtext}
\usepackage{tabularx,afterpage}
\usepackage[colorlinks=true
,urlcolor=blue
,anchorcolor=blue
,citecolor=blue
,filecolor=blue
,linkcolor=blue
,menucolor=blue
,linktocpage=true
,pdfproducer=medialab
,pdfa=true
]{hyperref}
\usepackage[utf8]{inputenc}
\usepackage{url}

\usepackage{csvsimple}
\usepackage[symbol]{footmisc}

\renewcommand{\arraystretch}{1.8}
\newcolumntype{L}[1]{>{\raggedright\let\newline\\\arraybackslash\hspace{0pt}}m{#1}}
\newcolumntype{C}[1]{>{\centering\let\newline\\\arraybackslash\hspace{0pt}}m{#1}}
\newcolumntype{R}[1]{>{\raggedleft\let\newline\\\arraybackslash\hspace{0pt}}m{#1}}

\newcommand {\be} {\begin {equation}}
\newcommand {\ee} {\end {equation}} 

\newcommand {\bes} {\begin {equation*}}
\newcommand {\ees} {\end {equation*}}

\newcommand{\beq}{\begin{equation}}
\newcommand{\eeq}{\end{equation}}
\newcommand{\bea}{\begin{eqnarray}}
\newcommand{\eea}{\end{eqnarray}}
\newcommand{\Eq}[1]{Eq.~(\ref{#1})}
\newcommand{\Eqs}[2]{Eqs.~(\ref{#1}) and (\ref{#2})}

\begin{document}

\title{Orbital Evolution of Satellite Galaxies in Self-Interacting Dark Matter Models}

\def\correspondingauthor{\thanks{Corresponding author - oslone@princeton.edu / os2124@nyu.edu}}

\author{Oren Slone}
\correspondingauthor{}
\affiliation{Department of Physics, Princeton University, Princeton, NJ 08544, USA}
\affiliation{Center for Cosmology and Particle Physics, Department of Physics, New York University, New York, NY 10003, USA}

\author{Fangzhou Jiang}
\affiliation{TAPIR, California Institute of Technology, Pasadena, CA 91125, USA}
\affiliation{Carnegie Observatories, 813 Santa Barbara Street, Pasadena, CA 91101, USA}

\author{Mariangela Lisanti}
\affiliation{Department of Physics, Princeton University, Princeton, NJ 08544, USA}

\author{Manoj Kaplinghat}
\affiliation{University of Irvine, Irvine, CA 92697, USA}


\begin{abstract}
Dark matter self interactions can leave distinctive signatures on the properties of satellite galaxies around Milky Way--like hosts through their impact on tidal stripping, ram pressure, and  gravothermal collapse. We delineate the regions of self-interacting dark matter parameter space---specified by interaction cross section and a velocity scale---where each of these effects dominates, and show how the relative mass loss depends on the satellite's initial mass, density profile and orbit. We obtain novel, conservative constraints in this parameter space using Milky Way satellite galaxies with notably high central densities and small pericenter distances. Our results for self-interacting dark matter models, in combination with constraints from clusters of galaxies, favor either velocity-dependent cross sections that lead to gravothermal core collapse in the densest satellites, or small cross sections which more closely resemble cold and collisionless dark matter.
\end{abstract}
\maketitle

\section{Introduction}
Current observational efforts are characterizing properties of satellite galaxies orbiting the Milky Way (MW)~\cite{2012AJ....144....4M, 2018A&A...619A.103F} and other hosts~\cite{Carlsten:2020fkn, Mao:2020rga}. This wealth of new data provides an exciting opportunity to test the evolution of small-scale structure in Self-Interacting Dark Matter~(SIDM)~\cite{Vogelsberger:2012ku,2013MNRAS.431L..20Z,2016MNRAS.461..710D,Read:2018pft,Robles:2019mfq,Zavala:2019sjk,Kaplinghat:2019svz,Kahlhoefer:2019oyt, Nadler:2020ulu,Turner:2020vlf, 2021MNRAS.503..920C,Sameie:2021ang,Sameie:2020xdi,Salucci:2018hqu}, as compared to Cold Dark Matter~(CDM)~\cite{2004ApJ...609..482K, 2010ApJ...709.1138D, 2010MNRAS.406.1290P, 2017MNRAS.471.4894D, 2017MNRAS.471.1709G, 2020MNRAS.491.1471S}.  SIDM arises generically in many theories~\cite{Tulin:2017ara}, and is well-motivated from an astrophysical standpoint~\cite{Kamada:2016euw,2017MNRAS.468.2283C,Ren:2018jpt, Kaplinghat:2019dhn}. In this study, we present a first analysis of the impact of velocity-dependent, anisotropic scattering between dark matter~(DM) particles on the orbits of satellite galaxies. 

For CDM, many ingredients affect satellite orbits including initial positions and velocities, as well as initial masses and concentrations, which affect mass loss due to tidal forces and orbital decay due to dynamical friction. For SIDM, these continue to play a role, while mass removal and momentum transfer from DM ram pressure also impact the evolution~\cite{Kummer:2017bhr}.  The density profile of an SIDM satellite is also critical to its evolution. The presence of a constant density, quasi-stable, isothermal core can increase the amount of mass that is tidally removed~\cite{2016MNRAS.461..710D}. This core can also undergo gravothermal collapse with a timescale that depends on various factors. High concentration satellites collapse faster~\cite{Kahlhoefer:2019oyt,Sameie:2019zfo}, with tidal stripping~\cite{Kaplinghat:2019svz} and high cross sections~\cite{Zavala:2019sjk} accelerating this process. The collapse can dramatically increase the central densities of satellites~\cite{Balberg:2002ue,2011MNRAS.415.1125K,Elbert:2014bma,Essig:2018pzq,Nishikawa:2019lsc,Kahlhoefer:2019oyt,Sameie:2019zfo,Turner:2020vlf}, which may be necessary for SIDM to be consistent with inferred densities of ultra-faint dwarfs~\cite{Zavala:2019sjk, Kahlhoefer:2019oyt,2021arXiv210609050K}.

We perform a conceptual analysis using semi-analytical orbit integration to map the consequences of SIDM on satellite orbital evolution. In particular, the parameter space for which mass loss in SIDM is dominated by either tidal stripping or ram-pressure evaporation is identified. In the former case, the signature is an overall mass reduction in the satellite's outer region, as compared to CDM. In the latter case, ram-pressure evaporation also removes mass from the satellite's central regions, causing lower central densities than would be expected for CDM. We also identify regions where the effects of gravothermal collapse are important. These findings motivate new observational handles for testing SIDM.

We begin with an overview of modeling satellite orbital evolution. We then discuss qualitative signatures of SIDM on satellite distributions around MW-like hosts. Building on this understanding, we set novel and conservative SIDM constraints using measured properties of MW dwarfs. The interested reader can find the precise details of our modeling in the Appendix and should note that effects not accounted for in our modeling do not change any qualitative results of the study and keep quantitive results conservative. We have compared the results of our modeling to results from the simulations of Ref.~\cite{Penarrubia:2010jk} and find qualitative agreement.

\section{Satellite Evolution}
This study considers scenarios where DM particles of mass $m_\chi$ interact via a light scalar/vector mediator of mass $m_\phi$ with coupling $\alpha_D$~\cite{Feng:2009mn, Loeb:2010gj,Kaplinghat:2015aga}. 
In the non-relativistic limit, elastic self-scattering is well-described by a Yukawa potential. When $\alpha_D m_\chi/m_\phi \ll 2\pi$, the cross section can be calculated using the Born approximation,
\beq
\frac{d\sigma}{d\theta} = \frac{\sigma_0 \sin{\theta}}{2 \left[1 + \frac{v^2}{\omega^2} \sin^2{\frac{\theta}{2}}\right]^2} \,\, ,
\label{eq:CS}
\eeq
where $\sigma_0 \equiv 4\pi \alpha_D^2 m_\chi^2/m_\phi^4$, $\omega \equiv m_\phi/m_\chi$, and $v$ is the velocity difference between interacting particles ($\sigma_{m}$ or $\sigma_{0m}$ will be used as shorthands for $\sigma/m_\chi$ or $\sigma_0/m_\chi$, respectively; in all equations we use the convention $c=1$.). Satellites are useful probes of Eq.~\eqref{eq:CS} because interactions between satellite and host probe velocities of order $\mathcal{O}(100)$~km/s while interactions within the satellite probe velocities of order $\mathcal{O}(10)$~km/s.

The three ingredients needed to properly model the evolution of in-falling satellites are: 

\vspace{0.05in}
\noindent
\emph{(i)} {\it Host \& Satellite Density Profiles.}--- Thermalization and heat flow resulting from self interactions alter DM density profiles~\cite{Spergel:1999mh}.  Any initially cuspy halo will exhibit inwards heat flow, resulting in a semi-stable isothermal configuration within the region where SIDM interactions are rapid. When baryons are unimportant, as is typically the case for low-mass satellites, the density profile develops a core at a radius $r_c$ and transitions to an NFW profile~\cite{1997ApJ...490..493N} beyond. We model the enclosed mass of such a profile as,
\beq
M_{\rm SIDM}(r) = M_{\rm NFW}(r) \cdot \text{tanh}\left(\frac{r}{r_c}\right) \, ,
\label{eq:Msidm}
\eeq
where $M_{\rm NFW}(r)$ is the enclosed mass of an NFW profile (with normalization $\rho_0$ and scale radius $r_s$), the core radius is defined as $r_c = \text{min}\left[0.5r_1, r_s \right]$ and $r_1$ is the radius below which interactions occur at least once within the age of the halo (see App.~\ref{sec:Model} for details). This form of the enclosed mass has an analytical solution for the density profile (found by differentiating Eq.~\eqref{eq:Msidm}) and also has the same total enclosed mass at $r \gg r_c$ as that of an NFW profile with the same $\rho_0$ and $r_s$. For a MW-like host, the presence of baryons leads to a cuspy profile that resembles an NFW profile with scale radius $r_s$~\cite{2014PhRvL.113b1302K, Sameie:2018chj}. We model such profiles, as well as profiles for a CDM model, as NFW profiles. The results of this study do not include the effects of the MW's stellar disk on subhalos~\cite{2010ApJ...709.1138D,2017MNRAS.471.1709G,2019MNRAS.487.4409K,2020MNRAS.491.1471S,2021MNRAS.503.4075G}, which requires a more sophisticated treatment. We can include the impact of an additional spherical potential (to partially mimic the stellar disk for the assumed orbits), but we find that its effects are small---see additional discussion below and in App.~\ref{sec:Sup_Figs}. We expect that the bound derived in Sec.~\ref{subsec:Constraints} will be more stringent when all the relevant physics is included.

 
This semi-stable configuration with an isothermal core does not remain intact indefinitely. Eventually, heat flow changes direction, transferring heat outwards. Because the core has negative specific heat, this results in a runaway gravothermal collapse process whereby the core heats up and shrinks simultaneously~\cite{Balberg:2002ue}. An important distinction should be made between the case when the SIDM Mean Free Path is either longer~(LMFP) or shorter~(SMFP) than the Jeans length of the core. In the LMFP regime, heat escapes efficiently from the core and the timescale for gravothermal collapse is $t_{\rm GC} \approx 290/(\langle \sigma_m \, v \rangle \, \rho_{\rm core})$, where $\rho_{\rm core}$ is the core density~\cite{Balberg:2002ue} (tidal stripping can significantly shorten $t_{\rm GC}$ by enhancing the rate of heat flow~\cite{Nishikawa:2019lsc,Kahlhoefer:2019oyt}). In this phase, the density profile has a shrinking core radius, $r_c$. Outside $r_c$, the density decreases approximately as $r^{-2.19}$ and transitions to the NFW profile when $r \gtrsim r_s$~\cite{Balberg:2002ue} for isolated halos. In the SMFP regime, heat gets trapped in the core and collapse occurs much more rapidly. Importantly, gravothermal collapse in the LMFP regime cannot produce arbitrarily large densities at a given radius because the evolution of the profile is such that the core density increases while the core size shrinks. To model the scenario of a halo that initially thermalizes and cores below $r_c$ and eventually undergoes collapse in the LMFP regime, we use a phenomenological profile of the form
\beq
\rho_{\rm GC}(r) = \frac{\rho_0}{(r/r_{\rm GC})^\alpha(1+r/r_{\rm GC})^{3-\alpha}} \,  \text{tanh}\left(\frac{r}{\text{min}[r_c,r_{\rm GC}]}\right)^{\alpha} \, ,
\label{eq:rho_GC}
\eeq
where $\alpha=2.19$ and $r_{\rm GC}=2r_s$ (see App.~\ref{sec:Model} for details).

\vspace{0.05in}
\noindent
\emph{(ii)} {\it Mass loss.}--- Satellite mass loss occurs predominantly via tidal stripping and ram-pressure evaporation. The former is a result of host tidal forces that strip material from the satellite, located at position $r$. These forces are efficient above the tidal radius, $\ell_t$ (measured from the satellite's center),
\beq
\ell_t \approx r \left(\frac{1}{g(r)} \frac{m_{\rm sat}(\ell_t)}{M_{\rm host}(r)}\right)^{1/3} \, ,
\label{eq:lt}
\eeq
where $m_{\rm sat}(\ell)$ and $M_{\rm host}(r)$ are the satellite and host enclosed masses respectively, and $g(r)$ accounts for both the tidal force and centrifugal
force acting on the satellite, as described in Eq.~\eqref{eq:lt_appendix} of App.~\ref{sec:Model}~\cite{1957ApJ...125..451V, 1962AJ.....67..471K}. The approximate mass-loss rate resulting from tidal stripping is
\beq
\dot{m}_{\rm TS} \approx - \mathcal{A} \frac{m_{\rm sat}(>\ell_t)}{t_{\rm dyn}(r)} \, ,
\label{eq:TS}
\eeq
where $\mathcal{A} = 0.55$~\cite{10.1093/mnras/stab696}, $m_{\rm sat}(>l_t) \equiv m_{\rm sat} - m_{\rm sat}(\ell_t)$ and $t_{\rm dyn}(r)$ is the dynamical time~\cite{Jiang:2020rdj}. The tidal radius is strongly dependent on the satellite's density profile. Comparing a cored SIDM to an NFW profile, one finds that when $\ell_t \lesssim r_c$, the tidal radius shrinks rapidly for SIDM, resulting in more efficient tidal stripping compared to CDM. 

Ram-pressure evaporation is the result of host-satellite DM interactions. For any scattering event, if a particle's final velocity is larger than the escape velocity of the satellite, $v_{\rm esc}$, the particle evaporates. The resulting mass-loss rate is
\beq
\dot{m}_{\rm RPe} \approx - m_{\rm sat} \, \eta_e \, \sigma_m  \, v_{\rm sat} \, \rho_{\rm host} \, ,
\label{eq:RP}
\eeq
where $v_{\rm sat}$ is the satellite's velocity relative to the host, $\rho_{\rm host}$ is the DM density of the host at position $r$, and $\eta_e$ is the $\mathcal{O}(1)$ evaporation fraction~\cite{Kummer:2017bhr}. For cases of interest, most host-satellite scattering events result in evaporation and the evaporation fraction is just unity, $\eta_e \approx 1$.

\begin{figure*}[htbp]
  \centering
  \includegraphics[width=.499\textwidth]{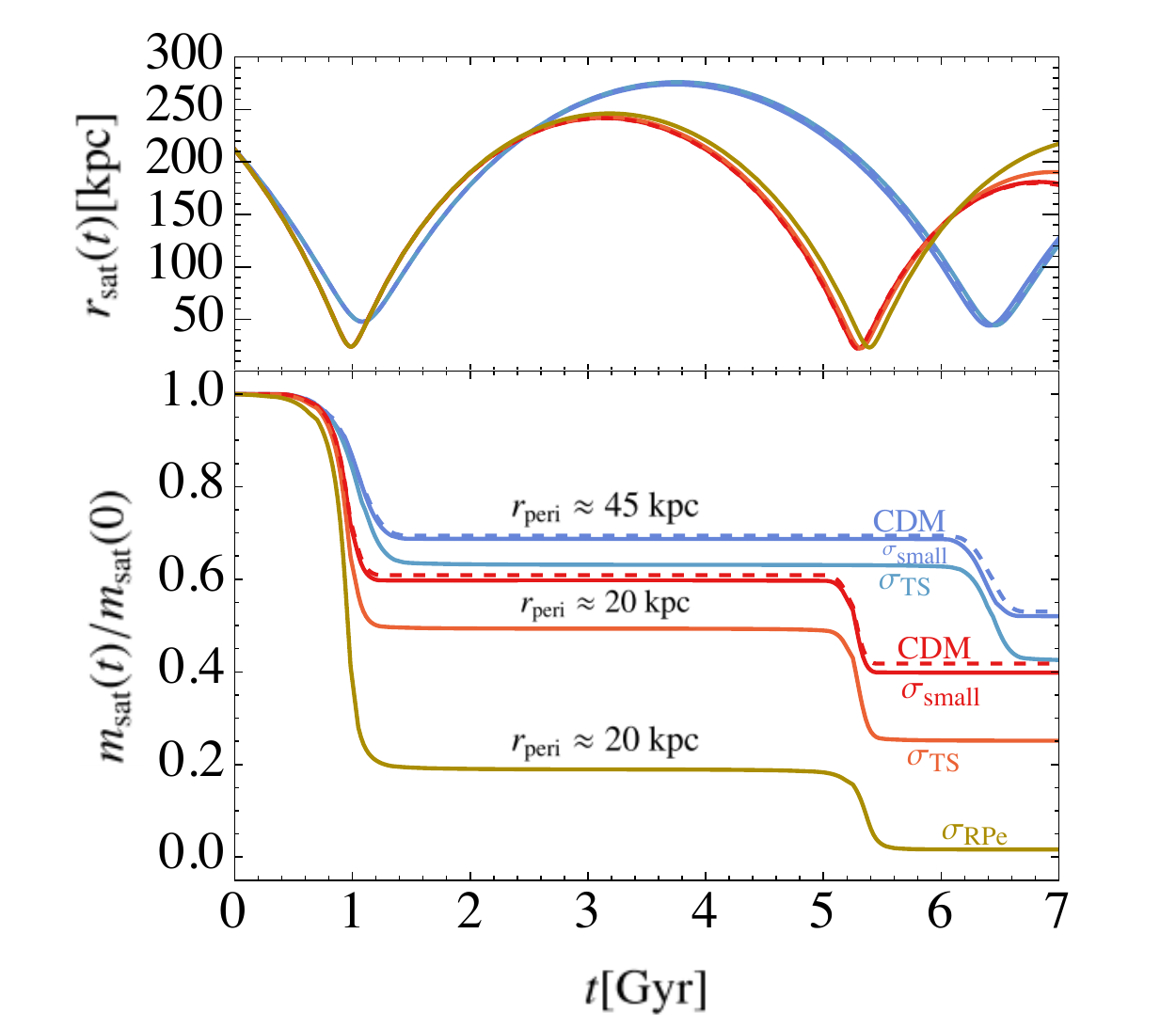}
  \includegraphics[width=.45\textwidth]{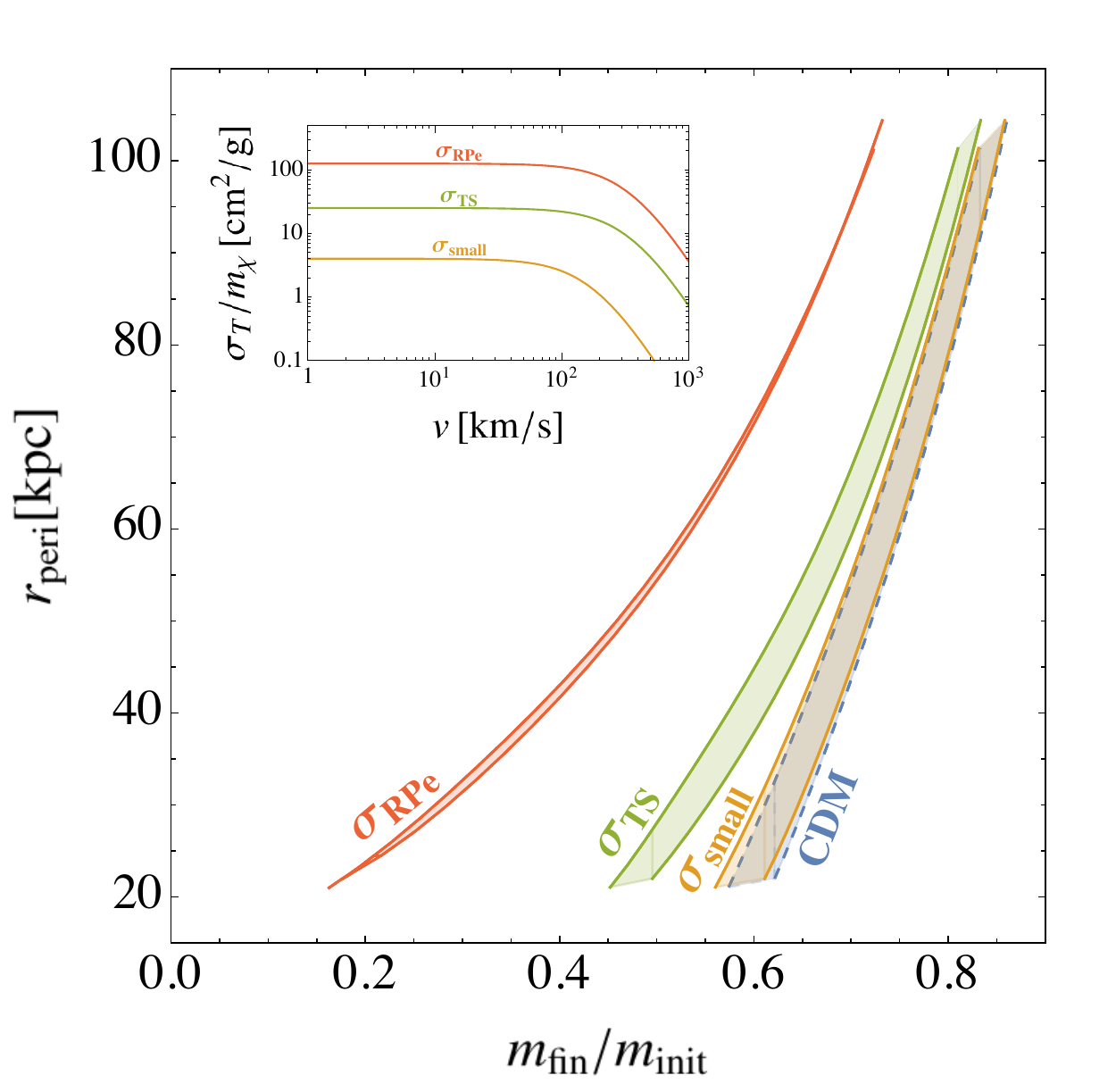}
  \caption{Satellite evolution around a $10^{12}~M_\odot$ host. \emph{(Left)} Time evolution of satellite mass \emph{(bottom) and radial position \emph{(top)}}, assuming $m_{\rm init} = 10^{10}\,M_\odot$, for CDM (dashed) and SIDM with initial isothermal cores (solid). The SIDM parameters $\{\sigma_{0m}, \omega\}$ correspond to: $\sigma_{\rm small} = \{4 \, {\rm cm}^2/{\rm g}, 160 \, {\rm km/s}\}$, $\sigma_{\rm TS} = \{25 \, {\rm cm}^2/{\rm g}, 315 \, {\rm km/s}\}$, and $\sigma_{\rm RPe} = \{125 \, {\rm cm}^2/{\rm g}, 315 \, {\rm km/s}\}$.  We contrast results for an eccentric orbit with pericenter $r_{\rm peri} \approx 20$~kpc with that of $r_{\rm peri} \approx 45$~kpc.  Relative to CDM, SIDM results in increased mass loss, especially in the case of more radial orbits. \emph{(Right)} Comparison of satellite mass loss in various scenarios. Initial satellite masses are considered in the range $m_{\rm init} \in [10^9, 10^{10.5}]~M_{\odot}$ corresponding to the width of each band. Satellites are placed on orbits with varying pericenters and evolved for one pericentric passage. The ratio of final to initial mass is given on the horizontal axis; the four bands correspond to CDM, $\sigma_{\rm small}$, $\sigma_{\rm TS}$ and $\sigma_{\rm RPe}$.  The inset shows the velocity dependence of the transfer cross section for each SIDM scenario.}
  \label{fig:mass_density}
\end{figure*}

\vspace{0.05in}
\noindent
\emph{(iii)} {\it Orbits.}--- Assuming that the satellite is not significantly deformed by tidal forces and treating it as a point, the orbit is obtained by evaluating the equation of motion,
\beq
\mathbf{a}_{\rm tot} = -\nabla \Phi + \mathbf{a}_{\rm DF} + \mathbf{a}_{\rm RPd} \, ,
\label{eq:EOM}
\eeq
where $\Phi$ is the gravitational potential of the host, $\mathbf{a}_{\rm DF}$ is the acceleration due to dynamical friction~\cite{1943ApJ....97..255C} (modeled using the Chadrasekhar formula), and $\mathbf{a}_{\rm RPd}$ is the acceleration due to
ram pressure. The latter arises from momentum transfer to the satellite from host-satellite SIDM scatterings, and is given by
\beq
 \mathbf{a}_{\rm RPd} \approx - \textbf{v} \, \eta_d \, \sigma_m \, v_{\rm sat} \, \rho_{\rm host} \, ,
\label{eq:RPd}
\eeq
where $\eta_d$ is the deceleration fraction~\cite{Kummer:2017bhr} (see Eq.~\eqref{eq:eta_d} and discussion around it). When $v_{\rm sat}$ is of order the virial velocity of the host, $\eta_d$ decreases rapidly as $\eta_d \propto (v_{\rm esc}/v_{\rm sat})^{2}$.  Thus, work done by ram-pressure deceleration is only comparable to dynamical friction for large $\sigma_{0m}$ and $\omega$ (see Fig.~\ref{fig:Rpd}). Full details are provided in App.~\ref{sec:Model}.

\section{Results}

\subsection{Orbit-Mass Relations}

Fig.~\ref{fig:mass_density} (bottom left) demonstrates the mass evolution of a $10^{10}~M_\odot$ satellite for different DM model assumptions. The satellite is initialized at the virial radius and virial velocity of a $10^{12}~M_\odot$ host at $z=1$, and two classes of orbits (with differing initial angular momenta) are considered with pericenters at $r_{\rm peri} \approx 20$ and 45~kpc (the top panel shows the satellites' radial positions, $r_{\rm sat}$, as functions of time). Dashed curves show results for CDM, while solid curves show results for three different sets of SIDM parameters, denoted as $\sigma_{\rm small}$, $\sigma_{\rm TS}$, and $\sigma_{\rm RPe}$ (defined in the caption).  $\sigma_{\rm small}$ gives results that are similar to CDM.  For $\sigma_{\rm TS}$, tidal stripping effects are more important for SIDM than for CDM. For $\sigma_{\rm RPe}$, the ram-pressure evaporation rate strongly exceeds that of tidal stripping.  Generally, satellite mass loss is more pronounced for SIDM than CDM, especially for larger interaction cross sections and more eccentric orbits.

The example of $\sigma_{\rm RPe}$ is particularly noteworthy.  After its first pericentric passage, $\sim$80\% of the satellite's mass is removed. The primary difference relative to $\sigma_{\rm small}$ and $\sigma_{\rm TS}$ is that mass loss due to ram-pressure evaporation dominates over tidal stripping, enhancing even further the difference between CDM and SIDM. While the latter preferentially removes mass from the outermost regions of the galaxy, the former has a more noticeable impact on its inner regions. This suggests that one can use observations of dwarf galaxies with low pericenters and high central densities to constrain SIDM.

We estimate the effects of the MW's stellar disk on the tidal stripping rate by reproducing Fig.~\ref{fig:mass_density} (left) with the inclusion of a $10^{11} \, M_\odot$ point mass at the MW's center. The result is shown in Fig.~\ref{fig:mass_density_Mdisk}. For the three reference cross sections, the mass loss difference is of order 10\% or less of $m_{\rm sat}(0)$. We defer a detailed analysis of these effects to future work.

Fig.~\ref{fig:mass_density} (right) demonstrates in more detail how mass loss over one pericentric passage varies as a function of the SIDM parameters, the initial satellite mass, $m_\text{init}$, and the pericentric radius, $r_\text{peri}$. Such results can potentially be used to infer microscopic properties of DM from observations such as the distribution of satellites around their hosts. The four bands correspond to CDM~(dashed) and the three SIDM cross sections of the left panel~(solid).  From right to left across each band, the initial masses span $m_\text{init} \in [10^9, 10^{10.5}] \, M_{\odot}$.  The results show that tidal stripping is more effective at creating differences between SIDM and CDM for larger mass satellites with smaller pericenters. The reason for this is two-fold. First, the more massive a satellite, the more dynamical friction it experiences causing its pericenter to decrease. Second, more massive satellites have larger values of $r_c$, which further enhance tidal stripping. For $\sigma_{\rm RPe}$, a sizable difference is already observed for large pericenters and for small pericenters the difference is dramatic, with SIDM satellites losing large fractions of their mass.

\begin{figure}
 \centering
  \includegraphics[width=0.5\textwidth]{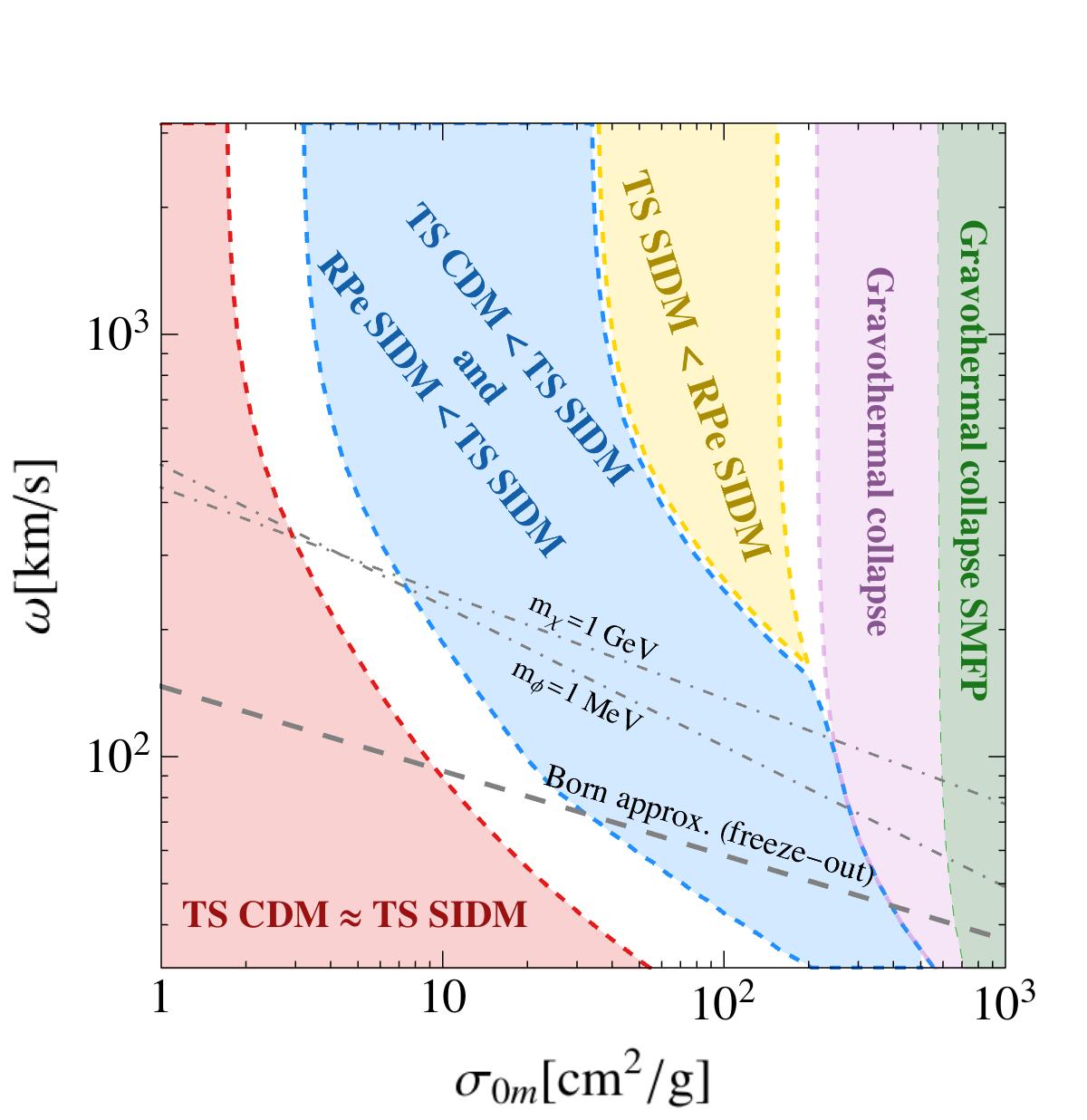}
  \caption{Schematic diagram showing regions in SIDM parameter space where particular effects dominate in affecting a satellite's mass-loss evolution within the initial mass range $m_{\rm init} \in [10^9, 10^{10}] \, M_{\odot}$ and for the orbital parameters as in the $r_{\rm peri} \approx 20$~kpc examples in Fig.~\ref{fig:mass_density}. White regions indicate parameters where not all satellites in the mass range fall into a given category. In the red region there is little observable difference in mass loss between CDM versus SIDM. In the blue region tidal stripping~(TS) is more effective in SIDM than in CDM, causing faster mass removal from the outskirts of SIDM satellites. In the yellow region ram-pressure evaporation~(RPe) dominates over tidal stripping for SIDM satellites, causing efficient mass removal from central regions. In the pink and green regions gravothermal collapse becomes relevant either in the long or short mean free path regimes for isolated satellites; to the left of these regions the collapse timescale can be reduced due to tidal stripping. Also shown in grey are curves relevant for a dark sector populated by thermal freeze-out of $\chi\chi \to \phi\phi$. Above the dashed curve the Born approximation holds for a scenario where $\alpha_D$ is fixed to give the correct dark matter thermal abundance. The dotted-dashed curves correspond to example values for $m_\chi$ and $m_\phi$.}
 \label{fig:parameter_space}
\end{figure}

\subsection{SIDM Parameter Space}

In Fig.~\ref{fig:parameter_space}, we build a conceptual map in SIDM parameter space showing where various mechanisms dominate a satellite's evolution. Colored areas demarcate regions of interest for satellites with $m_{\rm init} \in [10^9, 10^{10}]~M_{\odot}$ orbiting about a $10^{12}~M_\odot$ host. The concentration of each satellite corresponds to the best-fit concentration-mass relation of Ref.~\cite{2014MNRAS.441.3359D} at $z=1$ under the assumption that this relation approximately holds for SIDM cross sections considered in this study (see 
Ref.~\cite{2013MNRAS.430...81R} and discussion in the Appendix). The comparison between SIDM and CDM is done by averaging the mass loss over the last 2.5~Gyrs of 7 Gyr orbits.

In the red region, the ratio of mass loss between SIDM and CDM is less than 5\% and little difference is expected in satellite distributions. In the blue region, tidal stripping removes more mass from an SIDM satellite than from its CDM counterpart. Here, ram-pressure evaporation is small and therefore most mass is lost on the outskirts of the satellite. In the yellow region, ram-pressure evaporation in SIDM dominates over tidal stripping, and significant mass loss is expected from the centers of SIDM satellites.  These results differ from those in Ref.~\cite{Nadler:2020ulu}, which found that satellites are preferentially destroyed in SIDM halos solely because of ram pressure. This could potentially be explained  by the fact that Ref.~\cite{Nadler:2020ulu} used DM-only simulations with cored hosts (that have $\sim 35\%$ less mass below $\sim10$~kpc compared to their CDM counterpart), thereby reducing the effects of tidal stripping for satellites with small enough pericenters.

In the pink region, the gravothermal collapse timescale for an isolated halo is short, $t_{\rm GC}<10$~Gyrs. The green region within the pink denotes parameters for which the MFP is shorter than the Jeans length of the core and collapse occurs in the rapid SMFP regime. In both, tidal stripping and ram-pressure evaporation remain active, but calculating their effects requires accounting for the initially collapsed halo. A caveat to this is that even to the left of the pink region, where $t_{\rm GC}$ for an isolated halo is long, tidal stripping can significantly decrease its value. Therefore, the pink region should be thought of as parameters where LMFP gravothermal collapse can occur for isolated satellites before infall or with large pericenters.

A version of Fig.~\ref{fig:parameter_space} including the effects of a point mass of $10^{11}\,M_\odot$ (a proxy for the MW's stellar disk) is available in App.~\ref{sec:Sup_Figs}, Fig.~\ref{fig:parameter_space_Mdisk}. We find that the addition of this mass changes the results of the figure by a small amount, mostly increasing the area of the blue region and decreasing the area of the red region.

\begin{figure}
 \centering
  \includegraphics[width=0.5\textwidth]{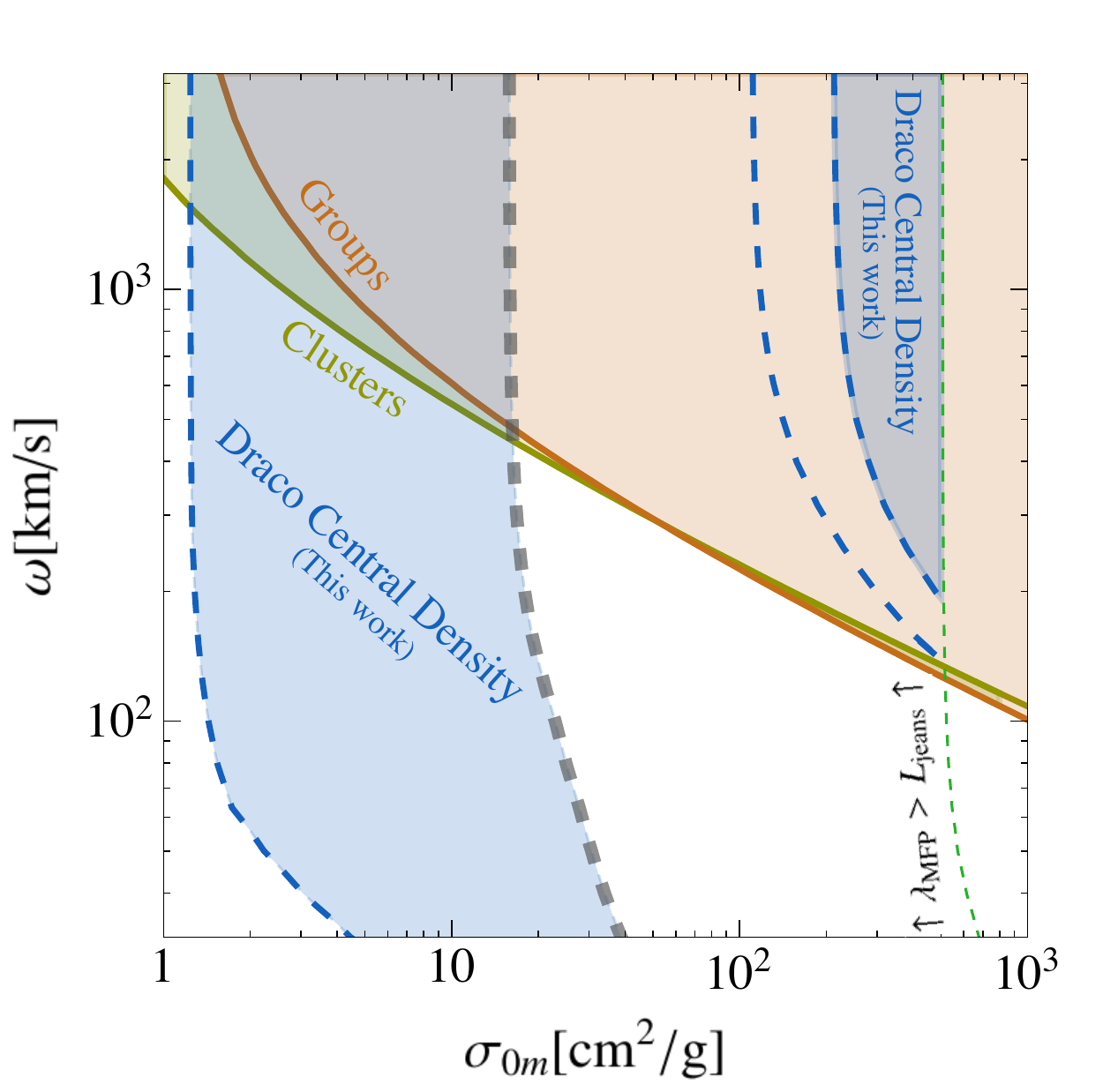}
  \caption{Conservative constraints based on measurements of Draco's central density, which arise from over-coring of an isothermal SIDM profile (left-most dark-blue region) and excessive mass removal from ram-pressure evaporation (right-most dark-blue region), shown for the $2\sigma$ lower limit on the measurement of $\rho_{150}$ for Draco.  Also shown are existing 95\% confidence-level bounds from galaxy groups (brown) and clusters (yellow)~\cite{Sagunski:2020spe}.  
}
 \label{fig:Draco}
\end{figure}

\subsection{Constraints from Dwarf Galaxy Measurements}
\label{subsec:Constraints}

\begin{figure*}[t]
 \centering
  \includegraphics[width=0.48\textwidth]{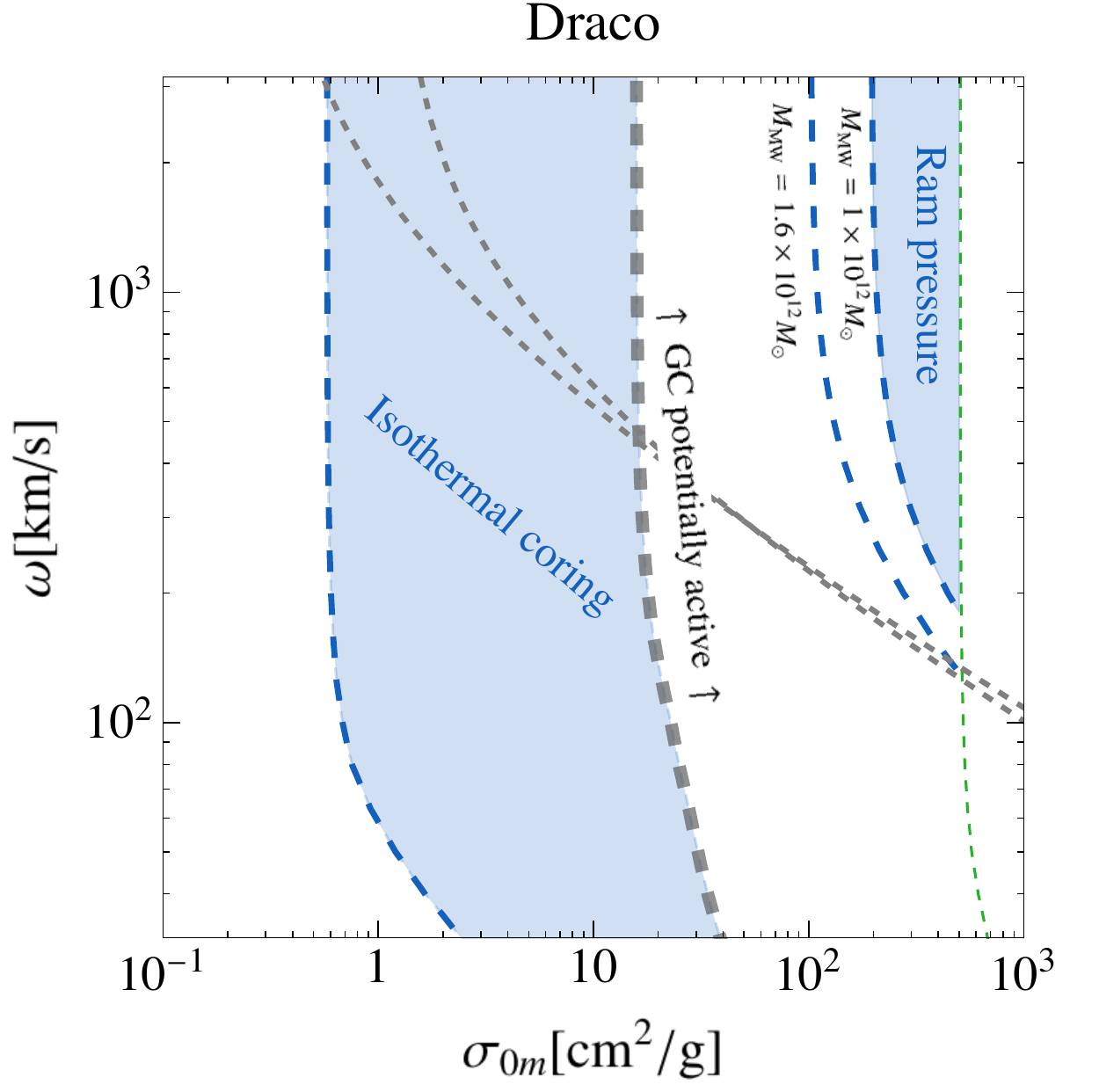}
  \includegraphics[width=0.48\textwidth]{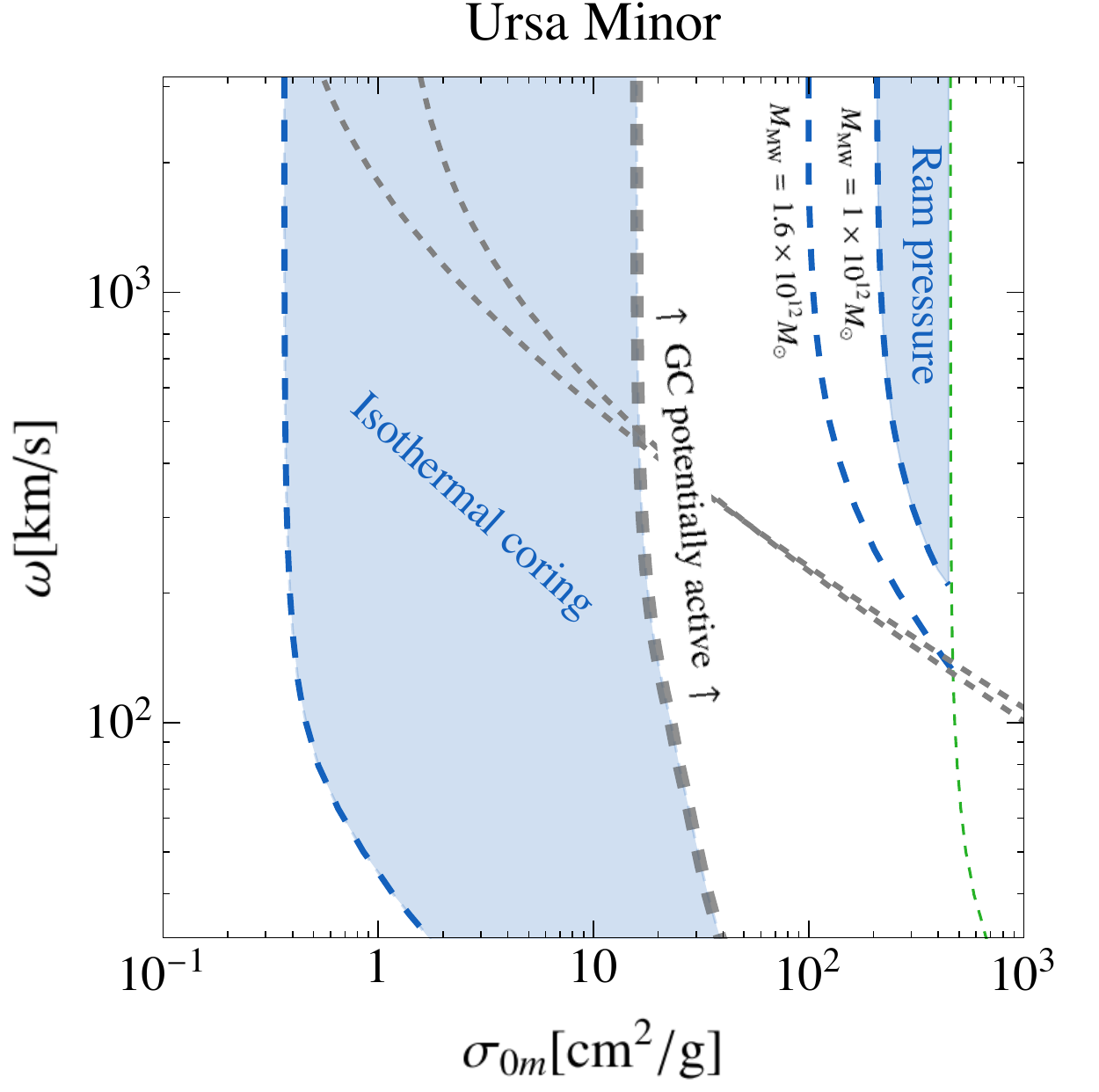} \\
  \vspace{0.1in}
  \includegraphics[width=0.48\textwidth]{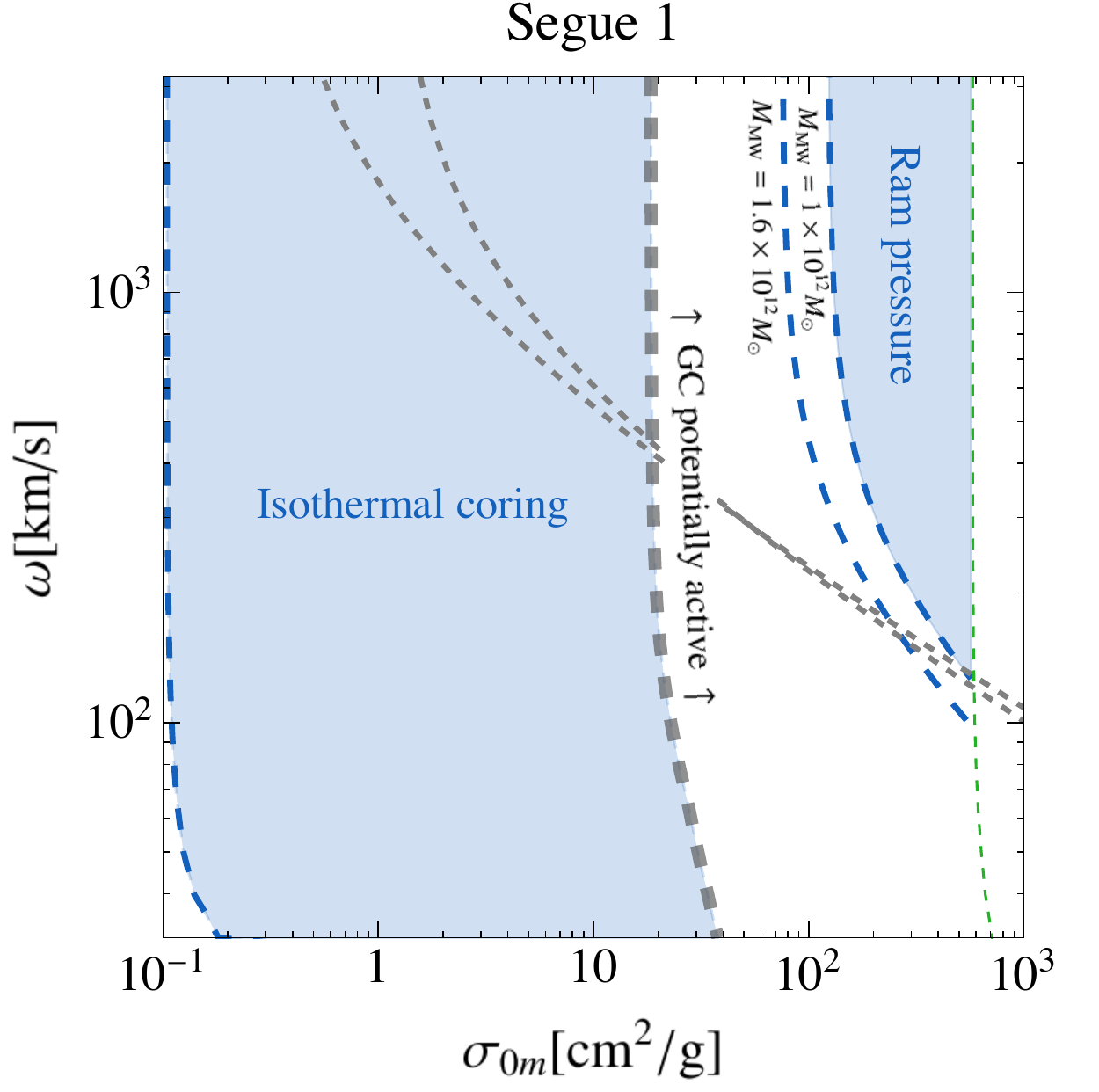}
  \includegraphics[width=0.48\textwidth]{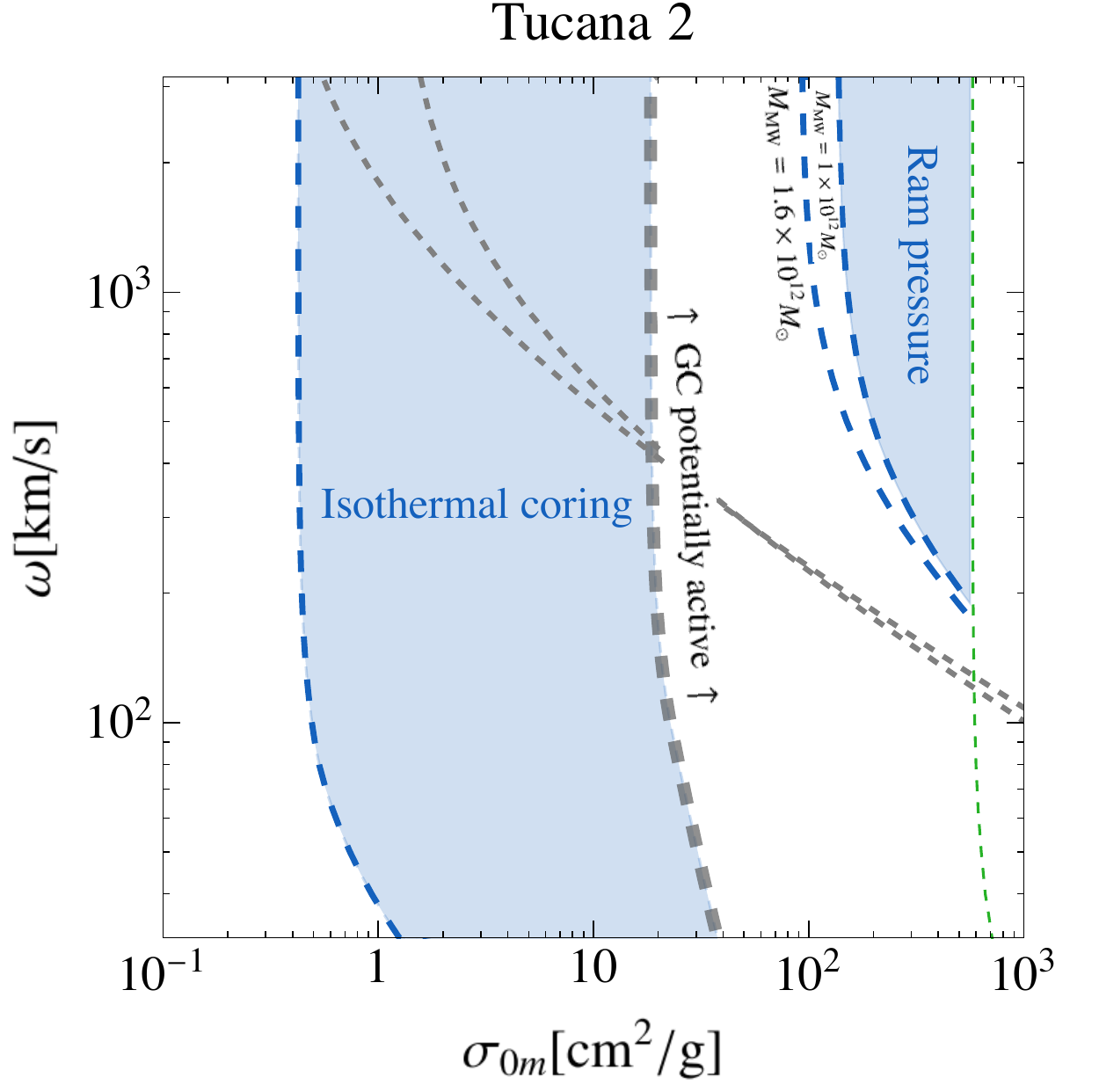}
  \caption{Contraints similar to those of Fig.~\ref{fig:Draco} for three additional systems: the classical dwarf Ursa Minor and the ultra-faint dwarfs Segue~1 and Tucana~2. For completeness, the Draco bound is shown here as well. For these contraints, the $1\sigma$ lower limit on measurements of $\rho_{150}$ or $\bar{\rho}_{1/2}$ are used.  Note that the Draco bound shown here takes the $1\sigma$ lower limit on the measured $\rho_{150}$, whereas that in Fig.~\ref{fig:Draco} takes the 2$\sigma$ lower limit. Also shown in dashed grey are existing 95\% confidence-level bounds from galaxy groups and clusters~\cite{Sagunski:2020spe}.}
  \label{fig:Other_Dwarfs}
\end{figure*}

A clear understanding of the orbital effects that dominate for different $\sigma_{0m}$ and $\omega$ is critical for testing SIDM. We provide a proof-of-concept example of how to perform such tests by using the central density of the Draco dwarf galaxy (Fig.~\ref{fig:Draco}), as well as those of Ursa Minor, Segue~1, and Tucana~2 (Fig.~\ref{fig:Other_Dwarfs}). These constraints rely on each dwarf's central density measurement which are provided in Table~\ref{tab:inputs}.

There are two constraints that can be obtained using dwarf galaxies with particularly high central densities. The first (``Isothermal-Coring Constraint'') arises when self interactions reduce the central density of the dwarf galaxy too much in a region of parameter space where gravothermal collapse is not possible. The second (``Ram-Pressure Constraint'') is a bound that arises from the potential of ram-pressure evaporation to remove too much mass from the interior of the dwarf galaxy. Precise details of how these constraints are conservatively placed are provided in App.~\ref{subsec:Other_Dwarfs}.

The first bound, shown as the left-most dark-blue region in Fig.~\ref{fig:Draco}, arises from the requirement that heating of the core not reduce Draco's present-day central density (at 150 pc), $\rho_{150}$, by more than $2 \sigma$ below its measured value~\cite{Kaplinghat:2019svz}. To obtain a conservative estimate of this bound, we marginalize over the unknown present-day mass of Draco by choosing the value that gives the largest central density at every point in parameter space. We also take the $2\sigma$ upper limit for Draco's concentration at $z=1$, which corresponds to the weakest constraint; we have verified that choosing a larger redshift produces a less conservative result. Finally, we require that $t_{\rm GC}$ is too long for gravothermal collapse to be active, even when tidal stripping effects are included. In the limit of constant interaction cross section, these results are consistent with Ref.~\cite{Read:2018pft}, albeit slightly weaker due to our conservative assumptions. A careful understanding of the role of gravothermal collapse in this region of parameter space enables us to extend the bounds to lower velocity scales compared to Ref.~\cite{Read:2018pft}.

The second bound, shown as the right-most dark-blue region in Fig.~\ref{fig:Draco}, arises in regimes where ram-pressure evaporation removes too much mass from the central regions of Draco during its most recent pericentric passage. Draco's orbit is calculated from its measured position and velocity~\cite{Patel_2020}. To conservatively estimate the bound, an initially fully gravothermally collapsed halo is taken, with a core radius that gives the maximal possible density at 150~pc, where Draco's density is measured. The inner (outer) dashed curves correspond to the $1\sigma$ lower (upper) limit on the MW mass, $M_{\rm MW} = (1.3 \pm 0.3) \times 10^{12} \, M_{\odot}$~\cite{Bland_Hawthorn_2016}. It should be noted that there are additional uncertainties (of similar order) associated with the orbit of Draco and other MW satellites~\cite{2018A&A...619A.103F,Patel_2020}. Better measurements of the orbital parameters would thus allow for a more robust constraint. We only evolve the satellite for one orbital period in order to avoid complications such as changes in the orbit induced by the LMC or by time dependence of the MW's potential; allowing for two full orbits results in stronger bounds that happen to closely track the outer dashed curve. It is likely that a full analysis will result in ruling out additional high $\sigma_{0m}$ regions, however our analysis also reveals that core collapse can insulate against that possibility. We note that the ram-pressure bound falls within a parameter space region that is anyway constrained by galaxy groups and clusters~\cite{Sagunski:2020spe}.

The conservative bounds presented in this study identify regions of SIDM parameter space where a dedicated analysis of all MW dwarfs should have excellent sensitivity. Indeed, we have derived bounds using Ursa Minor, Segue~1, and Tucana~2, which are all consistent with each other and shown in Fig.~\ref{fig:Other_Dwarfs}. For these results, we have used the $1\sigma$ lower limits on measurements of $\rho_{150}$ for Draco and Ursa Minor, and of $\bar{\rho}_{1/2}$ (the average central density within the half-light radius of the galaxy) for Segue~1 and Tucana~2. The choice of $1\sigma$, as opposed to the $2\sigma$ lower limit used for the Draco bound in the main text, illustrates the strong sensitivity of the isothermal coring bound to this choice. This sensitivity is due to the fact that central densities of an isothermal cored profile change slowly as $\sigma_{0m}$ is varied. The ram-pressure bounds are much less sensitive to small variations in the measurement of $\rho_{150}$ or $\bar{\rho}_{1/2}$.

Fig.~\ref{fig:Draco} also shows bounds from groups and clusters~\cite{Sagunski:2020spe}. There exist comparable bounds from oscillations of brightest cluster galaxies~\cite{Harvey:2018uwf} and a tighter bound from a strong lensing analysis in cluster galaxies~\cite{Andrade:2020lqq} with possible connections to core collapse~\cite{Meneghetti:2020yif,Minor:2020hic,Yang:2021kdf}. Combined with the cluster and group bounds, our results favor either a velocity-dependent SIDM cross sections that can trigger core collapse or small cross sections which more closely resemble CDM.

\section{Conclusions}
This study provides quantitative and intuitive understanding of the SIDM physics that affects internal properties and distributions of satellites within a MW--like host. The orbits of satellites with given initial masses, concentrations, and eccentricities were compared for different velocity-dependent, anisotropic cross sections. This semi-analytical approach is beneficial for identifying the physical mechanisms that affect satellite orbits in different regions of SIDM parameter space, and should hold so long as satellites are not significantly deformed along their orbits.

We identified several key regimes of interest in SIDM parameter space where particular mechanisms affect satellite orbits. Additionally, we placed conservative constraints on SIDM parameter space using measured central densities of Draco, Ursa Minor, Segue~1 and Tucana~2, which all provided consistent results. When combined with existing constraints, our findings strongly argue for SIDM models where velocity-dependent interactions can trigger gravothermal collapse in at least some satellites. We find that models with small cross sections are also allowed by our restricted analysis of the MW satellites. However, these models are essentially collisionless and we would categorize them as CDM rather than SIDM.

The results of this work underscore how individual MW satellites can provide remarkable sensitivity to velocity-dependent DM self interactions.  We also anticipate that the mass-loss and gravothermal collapse mechanisms studied here may translate into potentially observable differences in the population statistics of satellites.  With the abundance of observational data rapidly becoming available for both the MW and other MW--like systems in the Local Group, satellite galaxies will provide a critical---and potentially definitive---exploration of the SIDM parameter space.
\\\\
\section{Acknowledgements}

The authors gratefully acknowledge S.~Carlsten, J.~Greene, E.~Nadler, and P.~Natarajan for useful conversations. FJ is supported by the Troesh Scholarship at Caltech. MK is supported by the NSF under Award Number 1915005.  ML and OS are supported by the DOE under Award Number DE-SC0007968 and the Binational Science Foundation (grant~No.~2018140). ML is also supported by the Cottrell Scholar Program through the Research Corporation or  Science Advancement. OS is also supported by the NSF (grant~No.~PHY-1915409).
\vspace{0.4in}

\onecolumngrid
\appendix
\setcounter{equation}{0}
\setcounter{figure}{0}
\setcounter{table}{0}
\setcounter{section}{0}
\makeatletter
\renewcommand{\theequation}{S\arabic{equation}}
\renewcommand{\thefigure}{S\arabic{figure}}
\renewcommand{\thetable}{S\arabic{table}}
\makeatletter

\newpage

\begin{center}
\textbf{\large APPENDIX} \\
\vspace{0.05in}
\end{center}
\label{supp}

The first section of the Appendix provides a detailed description of the procedure used in this study to evolve the orbits of SIDM and CDM satellites. The second section describes the analysis performed to obtain the bounds in Fig.~\ref{fig:Draco} and Fig.~\ref{fig:Other_Dwarfs}. The third section provides supplementary figures that provide a preliminary exploration of the potential effects of the MW's stellar disk on our results.

\section{Detailed Procedure for Satellite Orbit Modeling}
\label{sec:Model}

As discussed in the main text, there are three key ingredients required to model a satellite's orbit in both the CDM and SIDM cases. This Appendix explores these ingredients in detail, focusing on: \emph{(a)}~the density profiles of the satellite and host halos, \emph{(b)}~the formalism for describing satellite mass loss from tidal stripping and ram-pressure evaporation, and \emph{(c)}~the numerical method for solving the equation of motion of the satellite.

We note that the modeling described below is meant to capture only qualitative behavior of a satellite galaxy orbiting a massive host. In particular, details such as tidal tracks~\cite{2008AN....329..934P}, effects related to shocking~\cite{2003ApJ...584..541H} and effects of the host's stellar content~\cite{2010ApJ...709.1138D} are not incorporated in our modeling.  Additionally, our modeling does not account for mass evolution of the host under the assumption that this effect is small for the case of the Milky Way~\cite{2020ARA_A..58..205H}. While such effects could change quantitative results, the specific changes to the constraints of Fig.~\ref{fig:Draco} and Fig.~\ref{fig:Other_Dwarfs} serve to shift curves towards constraining more parameter space (or moving them by a negligible amount) and therefore our results remain conservative.

\subsection{Host \& Satellite Density Profiles}
\label{subsec:Densities}

The density profile for a CDM satellite or host, as well as for a baryon-dominated SIDM host, is taken to be an NFW profile~\cite{1997ApJ...490..493N}. The enclosed mass of this profile has the form
\beq
M_{\rm NFW}(r) = 4 \pi \rho_0 r_s^3 \left[ \ln{\left(\frac{r+r_s}{r_s}\right)} - \frac{r}{r+r_s} \right],
\label{eq:Mnfw}
\eeq
where $r_s$ is the scale radius and $\rho_0$ is a normalization density. SIDM satellite profiles that are not gravothermally collapsed are modeled as Eq.~\eqref{eq:Msidm}. The density distributions corresponding to \Eqs{eq:Mnfw}{eq:Msidm} can be found by differentiating the formula for the enclosed mass, and are initially truncated at the virial radius. For the SIDM profile, the core density is $ 3/2 \times \rho_0r_s/r_c$.

Once the virial mass, $M_\text{200}$, of the halo is specified, its virial radius is determined by requiring that the average halo density is $200$ times the critical density, $\rho_{\rm crit} (z)$. The NFW scale radius then follows from the concentration-mass relation in Ref.~\cite{2014MNRAS.441.3359D} (which also agrees with the relations found in Refs.~\cite{Neto:2007vq} and~\cite{Diemer:2018vmz}), with $\rho_0$ obtained by requiring that the enclosed mass at $r_\text{200}$ gives $M_\text{200}$. Although the concentration-mass relation was derived for CDM, we assume that this relation approximately holds, or that possible variations to it do not have a large effect for SIDM cross sections considered in this study. Below $\sim 1$ cm$^2$/g, this has been verified in simulations~\cite{2013MNRAS.430...81R}. At cross sections which are much larger (of order $30$ cm$^2$/g and above) but below values for which gravothermal collapse occurs, core sizes could potentially be modified by such effects. Unless otherwise specified, the concentration of the satellite galaxy is evaluated at redshift $z=1$, corresponding to the time of infall in our examples. Note that, for the host halo specifically, we take a concentration of $c_{200} = 10$ and assume a total mass of $10^{12}~M_\odot$ at time of satellite infall, unless otherwise specified.  For the case of an NFW profile, this procedure is all that is required to set the free parameters of the halo model.  For the SIDM profile, this procedure sets the properties of the NFW profile before self interactions heat the central regions of the halo and the core forms.  During this process, some DM will be pushed out from the innermost region of the halo.  We have verified using idealized $N$-body SIDM simulations and cosmological FIRE-SIDM simulations~\cite{Li_2020} that the original NFW profile provides a good description beyond $r_c$.

The value of the radius $r_1$ is approximated using the following equation:
\beq
\langle \sigma_m \, v \rangle \, \rho_{\rm SIDM}(r_1) \, t_{\rm age} = 1,
\label{eq:App_r1}
\eeq
where $t_{\rm age}$ is the age of the satellite.  The velocity-averaged transfer cross section $\langle \sigma_m \, v \rangle$ is given by,
\beq
\langle \sigma_m \, v \rangle = \frac{1}{m_\chi} \int f(\mathbf{v}_1) f(\mathbf{v}_2) \, v \, \frac{d \sigma}{d \theta} \, \left( 1- \cos\theta \right) \, d^3 \mathbf{v}_1 d^3 \mathbf{v}_2 \, d\theta \, ,
\label{eq:sigma_avg}
\eeq
where $\mathbf{v} \equiv \mathbf{v}_1 - \mathbf{v}_2$ is the relative velocity, and $\theta$ is the scattering angle. This can be simplified to a single integral over $v$. Note that the $1- \cos\theta$ weighting does not suppress contributions from scattering events where the two particles exchange velocities ($\theta\simeq \pi$), which would not change the halo density profile. To take this into account, other weights such as $\sin^2\theta$ (viscosity cross section)~\cite{Tulin:2013teo} or $(1-|\cos\theta|)$~\cite{Robertson:2016qef,Robertson:2018anx} have been proposed. Importantly, all variations should reproduce the same $\langle \sigma_m \, v \rangle$ for values of $\omega$ much larger than the typical velocity dispersion of the satellite (up to $\mathcal{O}(1)$ factors). Therefore, any such variations will have small effects on our results since satellites have typical dispersions of $\mathcal{O}(10)$ km/s while we have considered $\omega \gtrsim 30$ km/s. The $(1-|\cos\theta|)$ weighted cross section is about a factor of 2 smaller for $w$ larger than the dispersion of DM, while for small $w$ the differences are 20-30\%. The $\sin^2\theta$ weighted cross section is different from the $1-\cos(\theta)$ by 33\% or lesser, depending on the value of $w$. This $O(1)$ systematic, which should be kept in mind when interpreting our results, should be resolved in the future with more simulations covering a range of $w$ values.
  
In the equation above, $f(\mathbf{v})$ is the Maxwell-Boltzmann velocity distribution for the DM,
\beq
f(\mathbf{v}) = \left( \frac{3}{ 2\pi \sigma_{\rm v}^2} \right)^{3/2} e^{- 3 \mathbf{v}^2 /2 \sigma_{\rm v}^2} 
\eeq
and $\sigma_{\rm v}$ is the root-mean-square velocity dispersion.  Assuming an isotropic velocity distribution, then $\sigma_{\rm v}^2 = 3 \sigma_r^2$, where the radial dispersion $\sigma_r$ follows from the radial Jeans equation,
\beq
\sigma_{r}^2(r) = \frac{1}{\rho_{\rm SIDM}(r)} \int_r^\infty \frac{\rho_{\rm SIDM} (r') \, v^2(r')}{r'} dr' \, ,
\eeq
with $v^2(r) = G M_{\rm SIDM}(r)/r$.  
With $r_1$ obtained in the way detailed above, we find that using $r_c = 0.5 r_1$ in \Eq{eq:Msidm} provides an accurate fit to the SIDM profile obtained from isothermal Jeans modeling or from idealized SIDM $N$-body simulations. 
For cross sections of $\sigma_m =1$--20 cm$^2$/g and for $t_{\rm age} = 1$--10~Gyr, \Eq{eq:Msidm}  with $r_c = 0.5 r_1$ agrees with simulation results to percent level.
For larger cross sections that yield $r_1>r_s$, we find that setting $r_c = r_s$ provides better fits. 

For certain ranges of parameter space, gravothermal collapse can affect the density distribution of SIDM halos and \Eq{eq:Msidm} no longer suffices. Gravothermally collapsed profiles in the LMFP regime are modeled based on the results of numerically solving for the self-similar solution in Ref.~\cite{Balberg:2002ue}. In this regime, the solution for the gravothermally collapsing region and its surroundings takes the form of a flat core within a shrinking radius, $r_c$.  Above this radius, the density decreases as $\rho_{\rm GC} \propto r^{-2.19}$ before transitioning to $\rho_{\rm NFW}(r)$ above the radius $r \gtrsim r_{\rm GC}$. This scenario is modeled by the profile given in Eq.~\eqref{eq:rho_GC}. This profile decreases as $r^{-3}$ when $r \gg r_{\rm GC}$, as $r^{-2.19}$ when $r_c<r<r_{\rm GC}$ and flattens out when $r<r_{\rm c}$, as would be expected from a collapsing core in the LMFP regime. We have verified that this density model is qualitatively similar to what was found in SIDM simulations with collapsing halos~\cite{Sameie:2019zfo,Nishikawa:2019lsc}. Note, however, that Ref.~\cite{Turner:2020vlf} finds a steeper single power-law profile in the region outside the core of $r^{-3}$ for satellites deep in the core-collapse phase. Such a change to the profile would mildly effect the ram-pressure bounds in Fig.~\ref{fig:Draco} and Fig.~\ref{fig:Other_Dwarfs}.

\subsection{Mass Loss}
\label{subsec:Mass_Loss}

Satellite mass is lost in one of two ways. For both SIDM and CDM scenarios, tidal stripping can remove mass from the outskirts of the satellite. The tidal stripping mass-loss rate is given by Eq.~\eqref{eq:TS}. The tidal radius, $\ell_t$, is calculated with Eq.~\eqref{eq:lt}, with $g(r)$ now written explicitly~\cite{1957ApJ...125..451V, 1962AJ.....67..471K}, 
\beq
\ell_t \simeq r \left[\frac{m_{\rm sat}(\ell_t)/M_{\rm host}(r)}{2-\frac{d\ln M_{\rm host}}{d\ln r} + \frac{v^2_{\rm tan}(r)}{v^2_{\rm circ}(r)}}\right]^{1/3},
\label{eq:lt_appendix}
\eeq
where $v_{\rm tan}(r) = |\hat{r} \times \mathbf{v}_{\rm sat} |$ is the instantaneous tangential velocity of the satellite and $v_{\rm circ}(r) = \sqrt{G M_{\rm host}(r)/r^2}$ is its circular velocity.  The dynamical time is taken to be
\beq
t_{\rm dyn}(r) = \sqrt{\frac{3 \pi}{16 G \bar{\rho}_{\rm host}(r)}} \, ,
\eeq
where $\bar{\rho}_{\rm host}$ is the average density of the host within radius $r$,~\cite{Jiang:2020rdj}.

For SIDM satellites only, ram-pressure evaporation can remove mass from all regions of the satellite. Ram-pressure evaporation is calculated according to Eq.~\eqref{eq:RP}, with the evaporation fraction given by
\beq
\eta_e = \frac{1}{\sigma} \int_{\pi - \theta_{\rm crit}}^{\theta_{\rm crit}} \frac{d\sigma}{d\theta} \, d\theta 
\quad , \text{where} \,\,\, \theta_{\rm crit} = \arccos\left( \frac{x^2 -1}{x^2 + 1} \right) 
\label{eq:eta_e}
\eeq
and $x = \bar{v}_{\rm esc}/v_{\rm sat}$ with $\bar{v}_{\rm esc}$ the average escape velocity of the satellite. The cross section, $\sigma_m$, in Eq.~\eqref{eq:RP} is evaluated at a velocity equal to the satellite's velocity with respect to the host plus the average escape velocity of the satellite, $v = v_{\rm sat} + \bar{v}_{\rm esc}$~\cite{Kummer:2017bhr}.  The latter is a reasonable approximation because host-satellite scattering events typically occur close to the center of the satellite.  In general, $\bar{v}_{\rm esc}/v_{\rm sat} \ll 1$ for the cases of interest here, so this approximation is sufficient even for scatterings that occur in the outer regions of the satellite.

Knowing both the mass-loss rate from tidal stripping and ram-pressure evaporation through Eqs.~\eqref{eq:TS} and~\eqref{eq:RP}, a prescription can be established to track the total mass and density profile of the satellite along its orbit.  For any small time step, $\Delta t = t'-t$, the tidal stripping or ram-pressure evaporation mass loss can be evaluated by
\beq
\Delta m_{\rm TS/RPe} = \int_t^{t'} \dot{m}_{\rm TS/RPe} \, dt \, ,
\eeq
where $\Delta m_{\rm TS/RPe}$ are typically negative.

Any mass that is removed via tidal stripping is taken away from the outer-most region of the satellite halo. This is modeled by defining a truncation radius that evolves over time.  At a given time step, the truncation radius is taken to decrease from $r_{\rm trunc}$ to $r_{\rm trunc}'$ such that the mass enclosed between these radii is equal to $\Delta m_{\rm TS}$, namely,
\beq
M(\rho_0,r_{\rm trunc}') - M(\rho_0,r_{\rm trunc}) = \Delta m_{\rm TS}.
\eeq
In contrast, mass that is removed via ram-pressure evaporation is taken from all regions of the satellite by changing the normalization of its density profile. This prescription removes mass from regions of the halo in a fashion that is linearly proportional to the local density at any given point. The normalization is taken to change from $\rho_0$ to $\rho_0'$ such that the mass difference is equal to $\Delta m_{\rm RPe}$, namely,
\beq
M(\rho_0',r_{\rm trunc}) - M(\rho_0,r_{\rm trunc}) = \Delta m_{\rm RPe}.
\eeq
If the velocity of the satellite is anomalously small, then $\Delta m_{\rm RPe}$ can be positive, i.e., the satellite accretes mass.  Although this is possible, it is never the case for the scenarios considered in this study.

\subsection{Orbits}
\label{subsec:Evolution}

The evolution of the satellite's orbit is obtained by solving the equation of motion, Eq.~\eqref{eq:EOM}. The dynamical friction is modeled using the Chadrasekhar formula~\cite{1943ApJ....97..255C},
\beq
\mathbf{a}_{\rm DF} = - 4 \pi G^2 \, m_{\rm sat} \, \rho_{\rm host} \ln{\Lambda} \, F_{\rm v}\left( v_{\rm sat}\right) \, \frac{\mathbf{v}_{\rm sat}}{v_{\rm sat}^3} \, ,
\eeq
where the Coulomb logarithm is defined as $\ln{\Lambda} = \text{min}[s,1] \ln{(M_{\rm host} / m_{\rm sat})}$ and is calibrated to simulations~\cite{2010MNRAS.408.2201G,2021MNRAS.503.4075G}, with $s = (3 r + r_{s,{\rm host}}) / (r + r_{s,{\rm host}})$ and $r_{s, {\rm host}}$ the scale radius of the host~\cite{Jiang:2020rdj}.  Assuming an isotropic and Maxwellian host halo, then $F_{\rm v}\left(v_{\rm sat} \right) = \text{Erf}(y) - 2 y e^{-y^2}/\sqrt{\pi} $ with $y = v_{\rm sat}/(\sqrt{2} \sigma_{\rm r})$, where $\sigma_r$ is the radial velocity dispersion of the host.

\begin{figure*}[t]
 \centering
  \includegraphics[width=0.5\textwidth]{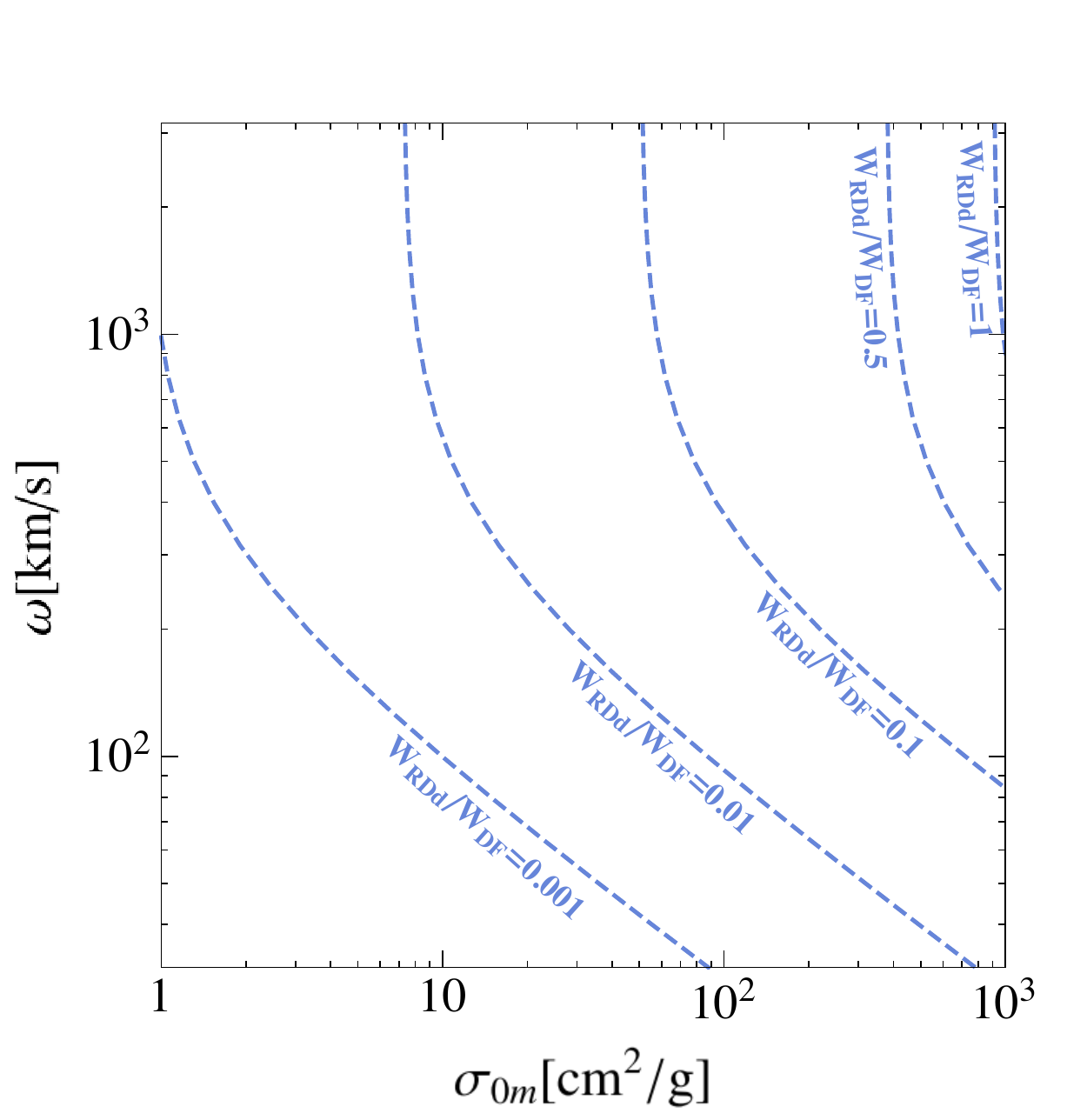}
  \caption{Contours of the ratio of work done by ram-pressure deceleration to the work done by dynamical friction over 7~Gyr orbits for $m_{\rm init} = 10^{10.5} \, M_{\odot}$ (the result is very weakly dependent on $m_{\rm init}$) and a $10^{12}~M_\odot$ host. In most of the parameter space, ram-pressure deceleration plays a subdominant role in the orbital evolution of satellites.}
  \label{fig:Rpd}
\end{figure*}

Ram-pressure deceleration, calculated according to Eq.~\eqref{eq:RPd}, can affect a satellite's orbit, especially in regions of large self-interaction cross sections. For $v_{\rm sat} \gg v_{\rm esc}$, the deceleration fraction, $\eta_d$, is
\beq
\eta_d = \frac{1}{m_\chi v \, \sigma} \int \Delta p_z \, \frac{d \sigma}{d \theta} \, d\theta \approx \frac{1}{2} \left( \frac{v_\text{esc}}{v_\text{sat}} \right)^{2}\, ,
\label{eq:eta_d}
\eeq
where $\Delta p_z$ is the change in momentum along the direction of motion of the incoming particles. The velocity dispersion, $v_{\rm disp}$, of interacting particles causes a further suppression of $\eta_d$ when $v_{\rm sat} \lesssim v_{\rm disp}$. However, at pericenter (when the effect is largest) the satellite's velocity is larger than the dispersions of both satellite and host, and the suppression does not enter the calculation. Figure~\ref{fig:Rpd} compares the role of ram-pressure deceleration to that of dynamical friction. The contours denote constant values of the ratio of work done by ram pressure to that done by dynamical friction, $W_{\rm RPd}/W_{\rm DF}$, over 7 Gyr orbits for a $m_{\rm init} = 10^{10.5} \, M_\odot$ satellite orbiting a $10^{12}~M_\odot$ host.  For most of the parameter space considered, orbital decay from ram-pressure deceleration is highly subdominant to the effects of dynamical friction and only for the largest cross sections considered do the two forces produce comparable work. Although the result is plotted only for a single satellite mass in Fig.~\ref{fig:Rpd}, the ratio $W_{\rm RPd}/W_{\rm DF}$ is only mildly dependent on $m_{\rm init}$. This occurs because of a cancellation between the explicit mass dependence of the dynamical friction force, and the mass dependence of the satellite's escape velocity, which enters the ram-pressure calculation.

To evaluate the equations of motions, \Eq{eq:EOM}, one must specify the initial conditions of the problem.  In our case, these are: a) the virial mass of the host halo, which sets the NFW density profile through the concentration relation, and b) the initial density profile, position and velocity of the satellite. Additionally, one must specify the SIDM parameters $\{\sigma_{0m},\omega\}$.  The density profile of the host galaxy is assumed to follow an NFW profile in all cases.  For the satellite, the initial profile depends on whether one is considering a CDM, isothermally cored SIDM, or a gravothermally collapsing scenario. For the case of CDM, one need only specify the mass of the halo, the scale radius is then set through the concentration-mass relation. For the isothermal cored SIDM scenario, after specifying the mass and calculating the scale radius, one must also specify $t_{\rm age}$, which sets $r_1$ through Eq.~\eqref{eq:App_r1}. In this study, we always take $t_{\rm age} = 10$~Gyrs; this assumption is only relevant when $r_c$ has not yet saturated $r_s$, which only occurs when $\sigma_{0m} \lesssim 20$ cm$^2/$g  for the range of masses considered in this study. 

After having specified these values, the evolution time is divided into small time steps such that $\Delta t \ll t_{\rm dyn}$ and slightly larger time-steps $\Delta T = (20\text{--}30) \Delta t$. At each $\Delta t$ interval, $\Delta m_{\rm TS}$ and $\Delta m_{\rm RPe}$ are calculated and $r_{\rm trunc}$ and $\rho_0$ are updated accordingly. The orbit is re-evaluated for every $\Delta T$ interval using \Eq{eq:EOM}, taking the masses, density profiles, positions and velocities from the end of the previous interval as the initial conditions. The larger $\Delta T$ interval is introduced for computational simplicity. We verify that the intervals are always small enough to have a negligible effect on the results.  The final result of the calculation is the mass, density profile, position and velocity of the satellite at all times. Additionally, one can sum over $\Delta m_{\rm TS}$ and $\Delta m_{\rm RPe}$ to evaluate the total mass lost to tidal stripping and ram-pressure evaporation and perform integrals over the dynamical friction and ram-pressure deceleration to calculate the work done by these forces.

\section{Constraints from Central Densities of Dwarf Galaxies}
\label{subsec:Other_Dwarfs}

This Appendix reviews in detail how to obtain conservative SIDM constraints using the central density of the Draco dwarf (Fig.~\ref{fig:Draco}), and the additional results for Ursa Minor, Segue~1, and Tucana~2 (Fig.~\ref{fig:Other_Dwarfs}). These constraints rely on present-day measurements of each dwarf's central density, which are provided in Table~\ref{tab:inputs}.  Note that for Draco and Ursa Minor, we use $\rho_{150}$, defined as the density at 150~pc.  For Segue~1 and Tucana~2, we use the average central density, $\bar{\rho}_{1/2}$, within $r_{1/2}$, the radius within which half of the galaxy's stellar luminosity is enclosed ($r_{1/2}=36$~pc for Segue~1~\cite{Martinez:2010xn} and $r_{1/2}=165$~pc for Tucana~2~\cite{Walker_2016}).

The goal of this procedure is to estimate the potential of dwarf observations in constraining SIDM parameter space.  Our approach is to make very conservative choices when it comes to assumptions that feed into the central density prediction.  As already highlighted in the main text, the results of this exercise demonstrate the important role played by Milky Way dwarfs in constraining $\sigma_{0m}$ and $\omega$.  This motivates pursuing a more rigorous likelihood procedure in future work.  For example, the conservative constraints presented in this work can be improved by performing a full Bayesian analysis that accounts for the unknown parameters with well-motivated priors, and appropriately stacks the contribution of each dwarf in the likelihood procedure.  

As discussed in the main text, there are two types of constraints that can be obtained: a `Ram-Pressure Constraint'' and an ``Isothermal-Coring Constraint''.

\vspace{0.1in}
\noindent
\textbf{Ram-Pressure Constraint}

\begin{figure}
  \centering
  \includegraphics[width=.99\textwidth]{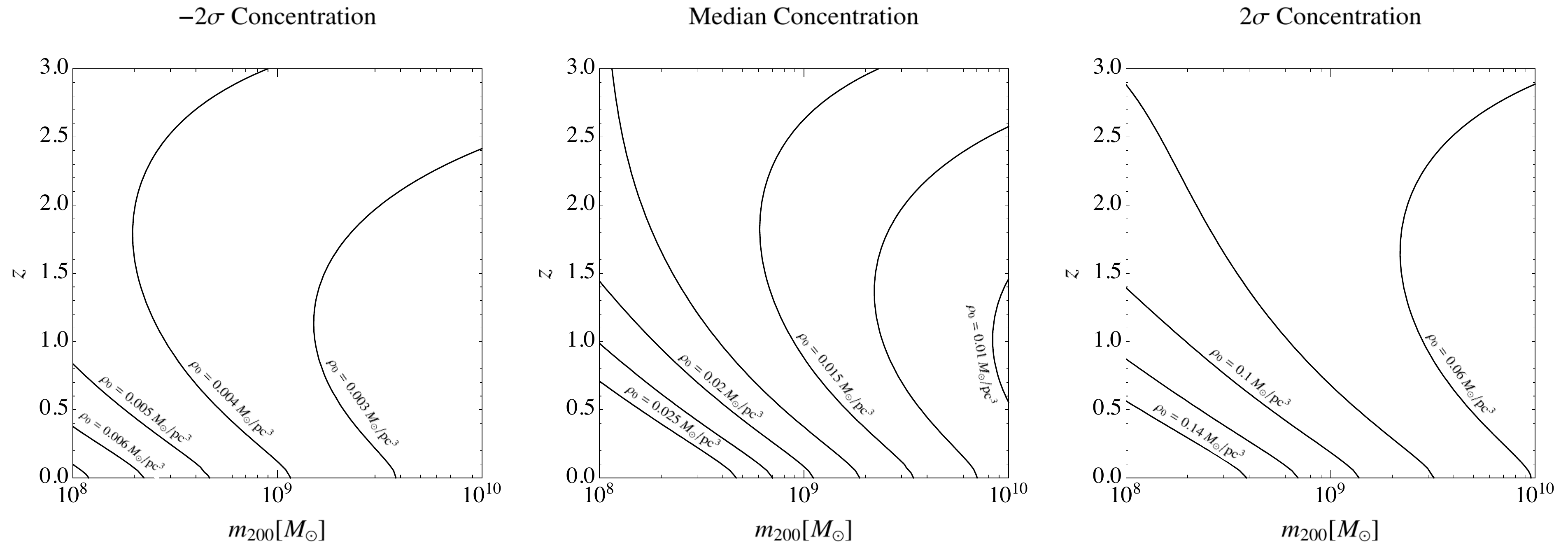}
  \caption{NFW profile scale densities as a function of $z$ and $m_{\rm 200}$ for varying distance from the median concentration-mass relation of Ref.~\cite{2014MNRAS.441.3359D}. In all panels, the central density increases, or is approximately constant, with decreasing $z$ for $z\lesssim2$. Note that $c_{200}$ is a function of both $z$ and $m_{200}$.}
  \label{fig:concentration_z}
\end{figure}

\vspace{0.1in}
\noindent To evaluate the impact of ram-pressure evaporation on any given dwarf, we infer its density and concentration at infall and initialize its energy and angular momentum given present-day observations of the system and the central density profile of the Milky Way. The orbit is then evolved forwards in time. Specifically:
\begin{enumerate}
\item The infall mass of each satellite, $M_{200}$, is provided in Table~\ref{tab:inputs}. For the classical dwarfs, Draco and Ursa Minor, the infall masses are taken from Ref.~\cite{Read:2018gpi}. These infall masses are inferred from abundance matching with mean star formation rates. For the ultra-faint dwarfs, Segue 1 and Tucana 2, the infall masses are taken to be $10^9\,M_{\odot}$, based on the approximate upper limit of high resolution cosmological simulations with low stellar content~\cite{Grand:2021fpx}. We assume here that infall properties and star formation physics are not significantly modified by self interactions, but this remains to be tested with simulations. 
\item The initial density distribution for the satellite is taken to be a fully gravothermally-collapsed profile following Eq.~\eqref{eq:rho_GC}. Importantly, gravothermal collapse in the LMFP regime cannot create arbitrarily large densities at any given radius or arbitrarily large average densities within any given radius. Specifically, below some minimal value, any additional decrease in the core size will not affect $\rho_{150}$ or $\bar{\rho}_{1/2}$ of the satellite. The value of $r_c$ is chosen such that this maximal central density is achieved for each satellite. For Draco and Ursa Minor, $r_c = 50$~pc, for Segue~1, $r_c = 2$ pc, and for Tucana~2, $r_c=20$ pc. These choices correspond to the largest possible initial conditions for the central density and are therefore maximally conservative.
\item The satellite's concentration, $c_{200}$, at time of infall is taken from the best-fit concentration-mass relation of Ref.~\cite{2014MNRAS.441.3359D}. We conservatively evaluate the concentration at $z=1$. The star formation histories of Draco, Ursa Minor,  Segue~1 and Tucana~2 suggest that their infall times may be closer to $z\sim 2$~\cite{2019arXiv190604180F,2021MNRAS.504.5270D}. We have verified that increasing the infall redshift decreases the initial central density of an NFW profile and would thus strengthen the constraints. Fig.~\ref{fig:concentration_z} shows this explicitly; we plot the NFW profile scale density for varying values of halo mass, redshift and distance from the median concentration-mass relation of Ref.~\cite{2014MNRAS.441.3359D}. We note that while concentration-mass relation of Ref.~\cite{2014MNRAS.441.3359D} agrees with Refs.~\cite{Neto:2007vq} and~\cite{Diemer:2018vmz}, it differs from Ref.~\cite{Ludlow:2016ifl}, and this is a potential caveat to the arguments made above.
\item The scale radius of the satellite's halo is determined from its concentration and $M_{200}$, which sets the virial radius, $r_{200}$, through the critical density, $\rho_{\rm crit}$.  Additionally, the overall normalization of the density distribution, $\rho_0$, is obtained by requiring that the enclosed mass at $r_{200}$ gives the virial mass.
\item The satellite's energy and angular momentum at $z=0$ is estimated using its present-day velocity and position and the density profile of the Milky Way. The observations are taken from Ref.~\cite{Patel_2020} for Draco, Ursa Minor, and Segue~1, and from Ref.~\cite{Simon_2018} for Tucana~2. The resulting values of pericenters for each of the satellites are $r_{\rm peri} \approx 44$ kpc for Draco, $r_{\rm peri} \approx 46$ kpc for Ursa Minor, $r_{\rm peri} \approx 19$ kpc for Segue~1 and $r_{\rm peri} \approx 38$ kpc for Tucana~2 which are all slightly higher than the median estimates of Ref.~\cite{2018A&A...619A.103F} but within the 2$\sigma$ error bands (smaller pericenters would correspond to slightly stronger constraints because ram-pressure evaporation would become more significant).
\item The satellite's orbit is obtained by placing it at some (arbitrary) initial position and evolving forwards in time for one pericentric passage, using the approximate values for energy and angular momentum determined in the previous step.  This procedure assumes that the energy and angular momentum of the satellite at infall match its present-day values, and does not account for losses due to dynamical friction and ram-pressure deceleration. While these corrections are likely negligible for the $\mathcal{O}(10^9)~M_\odot$ halos considered here, the choice of one pericentric passage minimizes the potential impact of these approximations. Additionally, the choice of a single pericentric passage avoids issues related to gravothermal collapse potentially occurring between passages. The constraints would significantly strengthen the more pericentric passages are included, since more mass is removed from the satellite during each of these.
\item After evolving for a single orbit, the central density (either $\rho_{150}$ or $\bar{\rho}_{1/2}$) is found and compared to the $2 \sigma$ lower limit from observations for results shown in Fig.~\ref{fig:Draco} and to the $1\sigma$ lower limit for results shown in Fig.~\ref{fig:Other_Dwarfs}.  If the predicted value is larger than the observational lower limit, then the point in the \{$\sigma_{0m}, \omega$\} parameter space is excluded.
\item The calculation is performed for two different Milky Way masses, $1\times 10^{12}$ and $1.6\times10^{12}$~$M_\odot$, which correspond to the lower and upper $1\sigma$ limits quoted in Ref.~\cite{Bland_Hawthorn_2016}.
\end{enumerate}

\begin{table}[t]
\footnotesize
\begin{center}
\begin{tabular}{C{2.5cm} | C{2.5cm}C{2.5cm}C{2.5cm}}
  \Xhline{3\arrayrulewidth}
\textbf{Dwarf Galaxy} 	&  \textbf{Central Density}   &  $\boldsymbol{M_{200}}$ & $\boldsymbol{c_{200}(z=1)}$ \\
					&  [$10^7~M_\odot~\text{kpc}^{-3}$]   &  $[M_\odot]$ &   \\
 \hline
 					\renewcommand{\arraystretch}{2.5}
 Draco 		&   $16.65$~\cite{Kaplinghat:2019svz}	&  	$1.8 \times 10^9$~\cite{Read:2018gpi}	& 	$8.84 \pm 3.36$ \\
 Ursa Minor  	&   $19.80$~\cite{Kaplinghat:2019svz}	& 	$2.8 \times 10^9$~\cite{Read:2018gpi}	&	$8.56 \pm 3.25$ \\
 Segue~1  	&   $92.71$~\cite{Martinez:2010xn}		&   	$10^{9}$~\cite{Grand:2021fpx} 			& 	$9.24 \pm 3.52$ \\
  Tucana~2  	&   $21.20$~\cite{Walker_2016}		&  	 $10^{9}$~\cite{Grand:2021fpx}			& 	$9.24 \pm 3.52$ \\
  \Xhline{3\arrayrulewidth}
\end{tabular}
\end{center}
\caption{The central densities for Draco and Ursa Minor correspond to the $1\sigma$ lower limit on the measured $\rho_{150}$, and for Segue~1 and Tucana~2 correspond to the $1\sigma$ lower limit on the measured $\bar{\rho}_{1/2}$. Note that in the main text the Draco bound corresponds to the $2\sigma$ lower limit. Also note that for Draco and Ursa Minor, we use $\rho_{150}$, defined as the density at 150~pc.  For Segue~1 and Tucana~2, we use the average central density, $\bar{\rho}_{1/2}$, within $r_{1/2}$, the radius within which half of the galaxy's stellar luminosity is enclosed ($r_{1/2}=36$~pc for Segue~1~\cite{Martinez:2010xn} and $r_{1/2}=165$~pc for Tucana~2~\cite{Walker_2016}). We use the concentration-mass relation from Ref.~\cite{2014MNRAS.441.3359D}. $M_{200}$ is the infall mass at $z=1$.  } 
\label{tab:inputs}
\end{table}

\vspace{0.1in}
\noindent
\textbf{Isothermal-Coring Constraint}

\vspace{0.1in}
\noindent To evaluate the impact of isothermal coring on any given dwarf, we determine its density distribution at present day while requiring that gravothermal collapse has not yet occurred.  Specifically:

\vspace{0.1in}
\begin{enumerate}
\item The density profile of the satellite is modeled by Eq.~\eqref{eq:Msidm}.  The halo is assumed to relax for $t_{\rm age} = 10$~Gyrs, which sets the value of $r_1$ following Eq.~\eqref{eq:App_r1}.
\item The central density of the profile depends on its current mass, $m_{\rm sat}$, and on the value of $r_s$. Because the current mass of the satellite is not well constrained, we evaluate the central density (either $\rho_{150}$ or $\bar{\rho}_{1/2}$) for a grid of masses. For a given value of $m_{\rm sat}$, the concentration of the satellite is taken from the $2\sigma$ upper limit of the concentration-mass relation from Ref.~\cite{2014MNRAS.441.3359D} at $z=1$. Then, for every point in the \{$\sigma_{0m}, \omega$\} parameter space, we choose the value of $m_{\rm sat}$ for which the central density is largest, i.e. the bound is weakest. We find that this value of $m_{\rm sat}$ is always less than the infall masses quoted in Table.~\ref{tab:inputs} and is around $10^{7.8 - 8.8} M_{\odot}$ for the four satellites considered in this study. These correspond to conservative choices for the satellite mass and concentration. The resulting value of the central density is then compared to the $2 \sigma$ lower limit  from observations for results shown in Fig.~\ref{fig:Draco} and to the $1\sigma$ lower limit for results shown in Fig.~\ref{fig:Other_Dwarfs}. If the predicted value is larger than the observational lower limit, then the point in the \{$\sigma_{0m}, \omega$\} parameter space is excluded.  This sets the left-most contour of the isothermal-coring bound. An important point is that as the cross section increases, the core size increases only until it reaches its maximal value at around $r_c \approx r_s$. For larger cross sections, the central density no longer decreases but rather remains constant. Therefore, much of the excluded parameter space is constrained at the same confidence level.
\item The right-most contour of the isothermal-coring bound is set by the requirement that gravothermal collapse be inactive.  Specifically, we estimate the gravothermal-collapse timescale, $t_{\rm GC}$, using~\cite{Balberg:2002ue}
\beq
t_{\rm GC} \approx \frac{290}{\langle \sigma_m v \rangle \rho_{\rm core}} \, ,
\eeq 
and require that $t_{\rm GC} > 20$~Gyrs. This choice accounts for potential shortening of $t_{\rm GC}$ through the effects of tidal stripping, based on the results of Ref.~\cite{Nishikawa:2019lsc}. For the velocity average, the dispersion is taken to be $\sigma_r = 1.1 \times v_{\rm max}/\sqrt{3}$, where $v_{\rm max}$ is the maximal circular velocity  of the NFW profile at radius $r_{\rm max}> r_c$. This may be derived from the maximum dispersion for a NFW halo ${\rm Max}(v_{\rm rms})$ using the Taylor-Navarro~\cite{Taylor:2001bq} phase-space density $Q(r)=\rho(r)/v_{\rm rms}(r)^3=0.3/(G v_{\rm max} r_{\rm max}^2 (r/r_s)^{-\eta}$ with $\eta\simeq 2$~\cite{Rocha:2012jg}. The median radial dispersion to $v_{\rm max}$ ratio plotted in Ref.~\cite{2008MNRAS.386.2022A} provides similar values (but about 10\% higher). In principle, the velocity average for this calculation should weight correctly for energy transfer and therefore has different powers of $v$ in Eq.~\eqref{eq:sigma_avg}~\cite{Colquhoun:2020adl}. However, for the reasons stated above, such variations to the calculation only change the result by $\mathcal{O}(1)$ for $\omega \gg \sigma_r$. We have verified that for $\omega \lesssim \sigma_r$, and with the assumption of weighting the cross section by $(v \sin{\theta})^2$, the value of $t_{\rm GC}$ increases such that constraints presented in this work are conservative.

In the time scale defined above, $\rho_{\rm core}$ is taken to be the central density of the isothermal cored SIDM profile, which should be interpreted as the minimum core density in the evolution of the halo. Note that since $v\propto v_{\rm max}$ and $\rho_{\rm core} \propto \rho_0$, this time scale has the same dependence on the initial profile as that in Ref.~\cite{Nishikawa:2019lsc}. However, the time scale used here is roughly a factor of two larger than that in  Ref.~\cite{Nishikawa:2019lsc}, which can be traced back to the choice of the LMFP conductivity normalization used to get a fit to halo profiles for moderate cross sections $\lesssim 10\rm\,cm^2/g$. The formula used here for $t_{\rm GC}$ is more consistent with the evolution of the core density for large cross sections~\cite{Balberg:2002ue,Essig:2018pzq}. We evaluate $t_{\rm GC}$ for a grid of masses in the range $m_{\rm sat} \in [10^{7}~M_{\odot}, M_{200}]$ for every point in the \{$\sigma_{0m}, \omega$\} parameter space. The constraints do not extend to points in parameter space where $t_{\rm GC} > 20$~Gyrs for any mass within this range.
\end{enumerate}

\noindent The final results for the classical dwarfs Draco and Ursa Minor, as well as the ultra-faint dwarfs Segue~1 and Tucana~2, are provided in Fig.~\ref{fig:Draco} and Fig.~\ref{fig:Other_Dwarfs}. The additional systems complement the result for Draco in a number of ways. Firstly, Draco's orbit has been shown to potentially be affected by the Large Magellenic Cloud (LMC)~\cite{Patel_2020}. If this is the case, then a full analysis should include the effects of Draco's interactions with the halo of the LMC and also account for the three-body orbit, both of which are beyond the scope of this work. However, the same study shows that Ursa Minor, which has a similar central density to Draco, is far less affected by the LMC. Specifically, the pericenter of Ursa Minor's orbit is expected to change far less when accounting for the multi-body orbit of the dwarf, LMC and the Milky Way. Second, Segue~1 and Tucana~2 are more DM dominated than either Draco or Ursa Minor, and thus have a different set of observational systematics. Finally, it is possible that the large central densities of the objects considered in this study could be the result of anomalously high concentrations, beyond even the $2 \sigma$ upper limit values used for results where gravothermal collapse does not occur ($2 \sigma$ was chosen specifically because we consider some of the densest known satellites of the Milky Way). If this is the case, the constraints would weaken. However, the combination of all four analyses illustrates the point that a future study of an ensemble of dwarfs will provide a robust constraint on the SIDM parameter space.

\section{Supplementary Figures}
\label{sec:Sup_Figs}

This section provides supplementary figures discussed in the main text. These figures quantify the effects of a spherical potential mimicking the MW's stellar disk on the mass removal rates of satellite galaxies considered in this study. In particular, we reproduce Fig.~\ref{fig:mass_density} (left) and Fig.~\ref{fig:parameter_space}, now including a point mass of $10^{11}~M_\odot$ (comfortably larger than the baryonic mass of the MW) at the MW's center. We find that for the orbits considered in this study, the results change by about 10\% or less of $m_{\rm sat}(0)$. The non-spherical nature of the disk and the associated effects of tidal heating and shocking are missing in this treatment. When these effects are included, we expect that the central densities of SIDM halos (that are not in the gravothermal core collapse phase) will be further lowered, thereby strengthening the bounds we have derived.\\\\

\begin{figure}
  \centering
  \includegraphics[width=.5\textwidth]{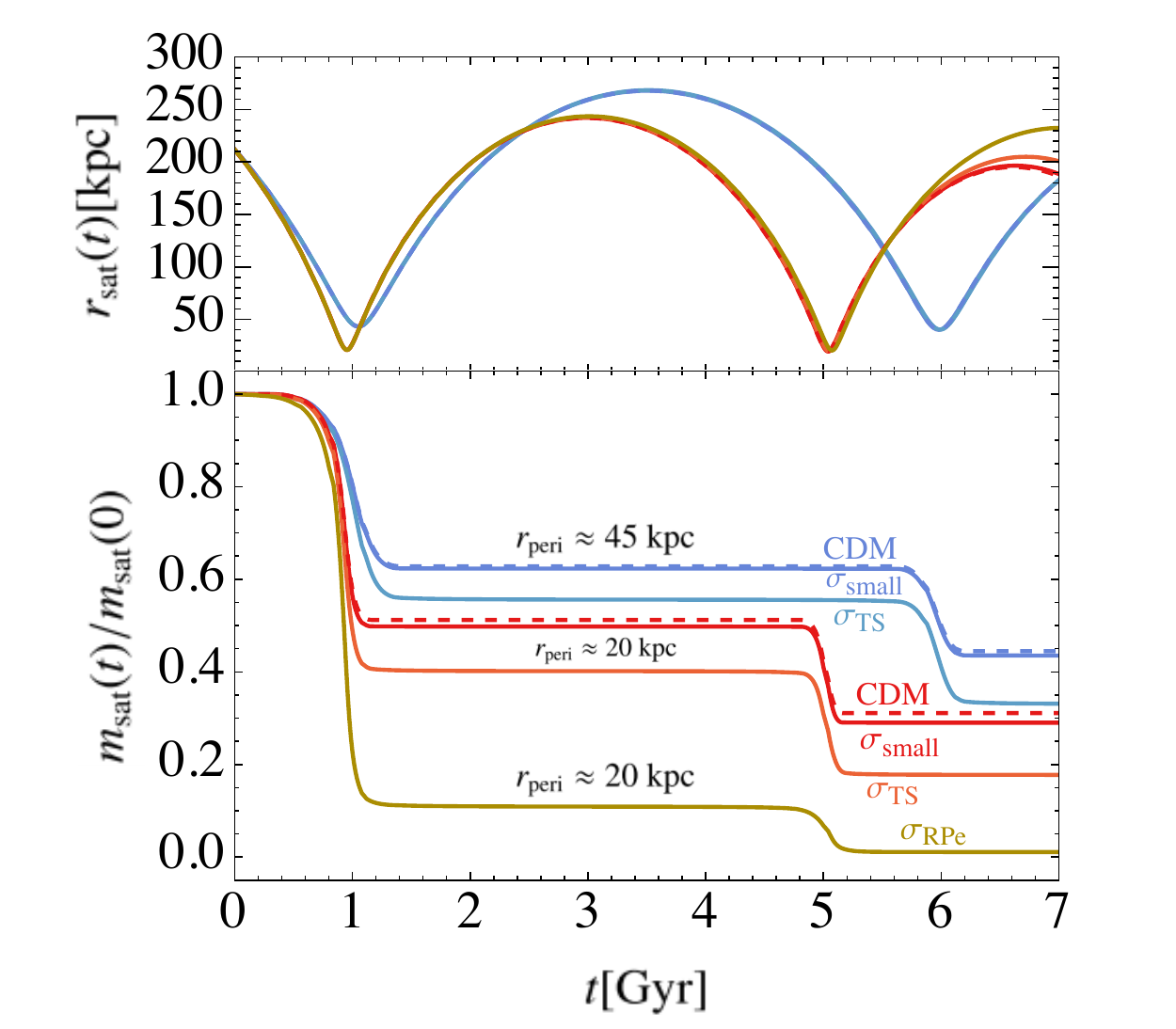}
  \caption{Same as Fig.~\ref{fig:mass_density} (left) but with the addition of a point mass of $10^{11}\,M_\odot$ at the MW's center.}
  \label{fig:mass_density_Mdisk}
\end{figure}

\begin{figure}
  \centering
  \includegraphics[width=.5\textwidth]{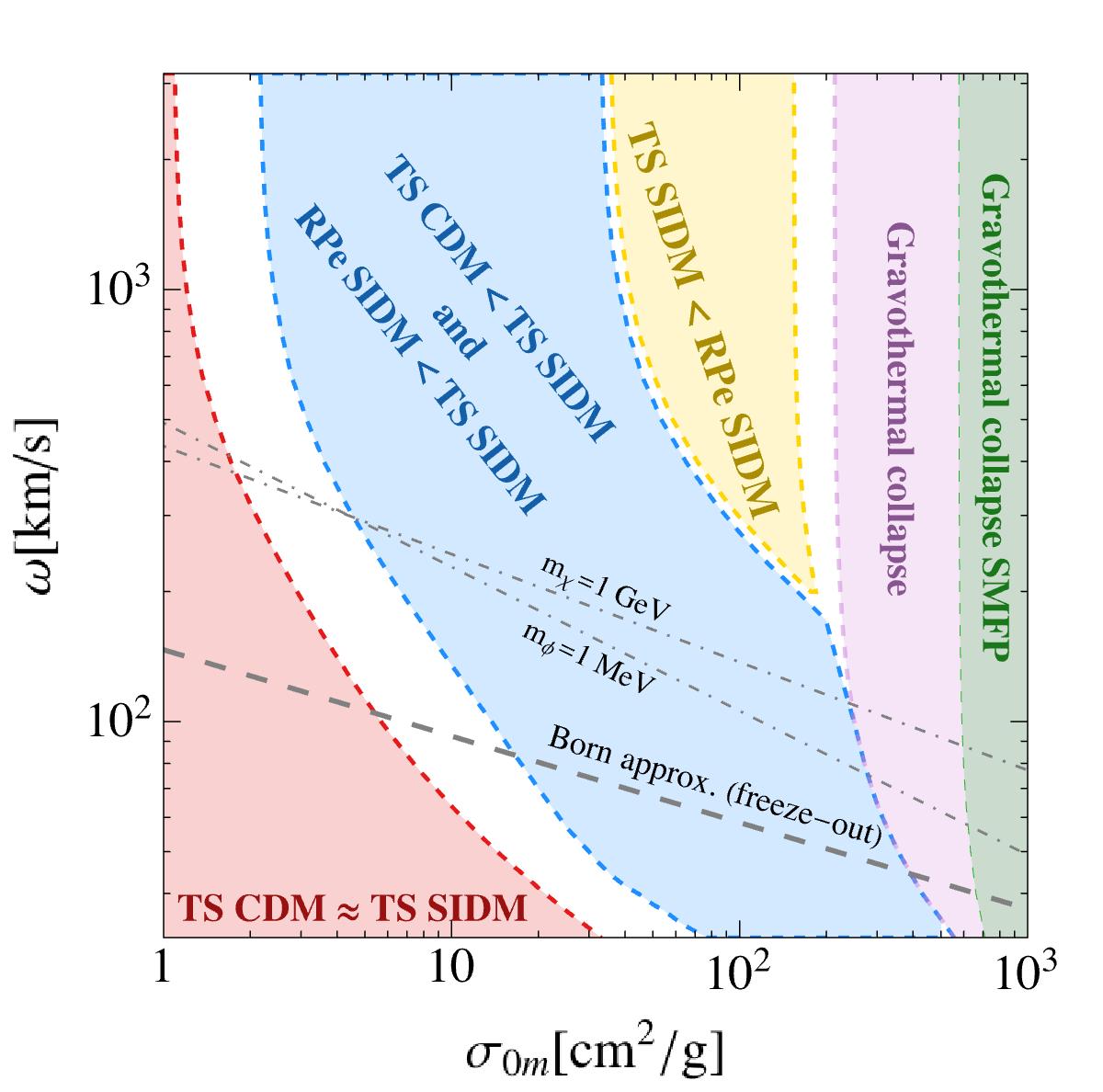}
  \caption{Same as Fig.~\ref{fig:parameter_space} but with the addition of a point mass of $10^{11}\,M_\odot$ at the MW's center.}
  \label{fig:parameter_space_Mdisk}
\end{figure}

\def\bibsection{} 
\bibliographystyle{apsrev}
\bibliography{sidm.bib}

\begin{thebibliography}{84}
\expandafter\ifx\csname natexlab\endcsname\relax\def\natexlab#1{#1}\fi
\expandafter\ifx\csname bibnamefont\endcsname\relax
  \def\bibnamefont#1{#1}\fi
\expandafter\ifx\csname bibfnamefont\endcsname\relax
  \def\bibfnamefont#1{#1}\fi
\expandafter\ifx\csname citenamefont\endcsname\relax
  \def\citenamefont#1{#1}\fi
\expandafter\ifx\csname url\endcsname\relax
  \def\url#1{\texttt{#1}}\fi
\expandafter\ifx\csname urlprefix\endcsname\relax\def\urlprefix{URL }\fi
\providecommand{\bibinfo}[2]{#2}
\providecommand{\eprint}[2][]{\url{#2}}

\bibitem[{\citenamefont{{McConnachie}}(2012)}]{2012AJ....144....4M}
\bibinfo{author}{\bibfnamefont{A.~W.} \bibnamefont{{McConnachie}}},
  \bibinfo{journal}{Astron. J.} \textbf{\bibinfo{volume}{144}},
  \bibinfo{eid}{4} (\bibinfo{year}{2012}), \eprint{1204.1562}.

\bibitem[{\citenamefont{{Fritz} et~al.}(2018)\citenamefont{{Fritz},
  {Battaglia}, {Pawlowski}, {Kallivayalil}, {van der Marel}, {Sohn}, {Brook},
  and {Besla}}}]{2018A&A...619A.103F}
\bibinfo{author}{\bibfnamefont{T.~K.} \bibnamefont{{Fritz}}},
  \bibinfo{author}{\bibfnamefont{G.}~\bibnamefont{{Battaglia}}},
  \bibinfo{author}{\bibfnamefont{M.~S.} \bibnamefont{{Pawlowski}}},
  \bibinfo{author}{\bibfnamefont{N.}~\bibnamefont{{Kallivayalil}}},
  \bibinfo{author}{\bibfnamefont{R.}~\bibnamefont{{van der Marel}}},
  \bibinfo{author}{\bibfnamefont{S.~T.} \bibnamefont{{Sohn}}},
  \bibinfo{author}{\bibfnamefont{C.}~\bibnamefont{{Brook}}}, \bibnamefont{and}
  \bibinfo{author}{\bibfnamefont{G.}~\bibnamefont{{Besla}}},
  \bibinfo{journal}{Astron. Astrophys.} \textbf{\bibinfo{volume}{619}},
  \bibinfo{eid}{A103} (\bibinfo{year}{2018}), \eprint{1805.00908}.

\bibitem[{\citenamefont{Carlsten et~al.}(2020)\citenamefont{Carlsten, Greene,
  Peter, Greco, and Beaton}}]{Carlsten:2020fkn}
\bibinfo{author}{\bibfnamefont{S.~G.} \bibnamefont{Carlsten}},
  \bibinfo{author}{\bibfnamefont{J.~E.} \bibnamefont{Greene}},
  \bibinfo{author}{\bibfnamefont{A.~H.~G.} \bibnamefont{Peter}},
  \bibinfo{author}{\bibfnamefont{J.~P.} \bibnamefont{Greco}}, \bibnamefont{and}
  \bibinfo{author}{\bibfnamefont{R.~L.} \bibnamefont{Beaton}},
  \bibinfo{journal}{Astrophys. J.} \textbf{\bibinfo{volume}{902}},
  \bibinfo{pages}{124} (\bibinfo{year}{2020}), \eprint{2006.02444}.

\bibitem[{\citenamefont{Mao et~al.}(2021)\citenamefont{Mao, Geha, Wechsler,
  Weiner, Tollerud, Nadler, and Kallivayalil}}]{Mao:2020rga}
\bibinfo{author}{\bibfnamefont{Y.-Y.} \bibnamefont{Mao}},
  \bibinfo{author}{\bibfnamefont{M.}~\bibnamefont{Geha}},
  \bibinfo{author}{\bibfnamefont{R.~H.} \bibnamefont{Wechsler}},
  \bibinfo{author}{\bibfnamefont{B.}~\bibnamefont{Weiner}},
  \bibinfo{author}{\bibfnamefont{E.~J.} \bibnamefont{Tollerud}},
  \bibinfo{author}{\bibfnamefont{E.~O.} \bibnamefont{Nadler}},
  \bibnamefont{and}
  \bibinfo{author}{\bibfnamefont{N.}~\bibnamefont{Kallivayalil}},
  \bibinfo{journal}{Astrophys. J.} \textbf{\bibinfo{volume}{907}},
  \bibinfo{pages}{85} (\bibinfo{year}{2021}), \eprint{2008.12783}.

\bibitem[{\citenamefont{Vogelsberger et~al.}(2012)\citenamefont{Vogelsberger,
  Zavala, and Loeb}}]{Vogelsberger:2012ku}
\bibinfo{author}{\bibfnamefont{M.}~\bibnamefont{Vogelsberger}},
  \bibinfo{author}{\bibfnamefont{J.}~\bibnamefont{Zavala}}, \bibnamefont{and}
  \bibinfo{author}{\bibfnamefont{A.}~\bibnamefont{Loeb}},
  \bibinfo{journal}{Mon. Not. Roy. Astron. Soc.}
  \textbf{\bibinfo{volume}{423}}, \bibinfo{pages}{3740} (\bibinfo{year}{2012}),
  \eprint{1201.5892}.

\bibitem[{\citenamefont{{Zavala} et~al.}(2013)\citenamefont{{Zavala},
  {Vogelsberger}, and {Walker}}}]{2013MNRAS.431L..20Z}
\bibinfo{author}{\bibfnamefont{J.}~\bibnamefont{{Zavala}}},
  \bibinfo{author}{\bibfnamefont{M.}~\bibnamefont{{Vogelsberger}}},
  \bibnamefont{and} \bibinfo{author}{\bibfnamefont{M.~G.}
  \bibnamefont{{Walker}}}, \bibinfo{journal}{Mon. Not. Roy. Astron. Soc}
  \textbf{\bibinfo{volume}{431}}, \bibinfo{pages}{L20} (\bibinfo{year}{2013}),
  \eprint{1211.6426}.

\bibitem[{\citenamefont{{Dooley} et~al.}(2016)\citenamefont{{Dooley}, {Peter},
  {Vogelsberger}, {Zavala}, and {Frebel}}}]{2016MNRAS.461..710D}
\bibinfo{author}{\bibfnamefont{G.~A.} \bibnamefont{{Dooley}}},
  \bibinfo{author}{\bibfnamefont{A.~H.~G.} \bibnamefont{{Peter}}},
  \bibinfo{author}{\bibfnamefont{M.}~\bibnamefont{{Vogelsberger}}},
  \bibinfo{author}{\bibfnamefont{J.}~\bibnamefont{{Zavala}}}, \bibnamefont{and}
  \bibinfo{author}{\bibfnamefont{A.}~\bibnamefont{{Frebel}}},
  \bibinfo{journal}{Mon.~Not.~Roy.~Astron.~Soc.}
  \textbf{\bibinfo{volume}{461}}, \bibinfo{pages}{710} (\bibinfo{year}{2016}),
  \eprint{1603.08919}.

\bibitem[{\citenamefont{Read et~al.}(2018)\citenamefont{Read, Walker, and
  Steger}}]{Read:2018pft}
\bibinfo{author}{\bibfnamefont{J.~I.} \bibnamefont{Read}},
  \bibinfo{author}{\bibfnamefont{M.~G.} \bibnamefont{Walker}},
  \bibnamefont{and} \bibinfo{author}{\bibfnamefont{P.}~\bibnamefont{Steger}},
  \bibinfo{journal}{Mon. Not. Roy. Astron. Soc.}
  \textbf{\bibinfo{volume}{481}}, \bibinfo{pages}{860} (\bibinfo{year}{2018}),
  \eprint{1805.06934}.

\bibitem[{\citenamefont{Robles et~al.}(2019)\citenamefont{Robles, Kelley,
  Bullock, and Kaplinghat}}]{Robles:2019mfq}
\bibinfo{author}{\bibfnamefont{V.~H.} \bibnamefont{Robles}},
  \bibinfo{author}{\bibfnamefont{T.}~\bibnamefont{Kelley}},
  \bibinfo{author}{\bibfnamefont{J.~S.} \bibnamefont{Bullock}},
  \bibnamefont{and}
  \bibinfo{author}{\bibfnamefont{M.}~\bibnamefont{Kaplinghat}},
  \bibinfo{journal}{Mon. Not. Roy. Astron. Soc.}
  \textbf{\bibinfo{volume}{490}}, \bibinfo{pages}{2117} (\bibinfo{year}{2019}),
  \eprint{1903.01469}.

\bibitem[{\citenamefont{Zavala et~al.}(2019)\citenamefont{Zavala, Lovell,
  Vogelsberger, and Burger}}]{Zavala:2019sjk}
\bibinfo{author}{\bibfnamefont{J.}~\bibnamefont{Zavala}},
  \bibinfo{author}{\bibfnamefont{M.~R.} \bibnamefont{Lovell}},
  \bibinfo{author}{\bibfnamefont{M.}~\bibnamefont{Vogelsberger}},
  \bibnamefont{and} \bibinfo{author}{\bibfnamefont{J.~D.}
  \bibnamefont{Burger}}, \bibinfo{journal}{Phys. Rev. D}
  \textbf{\bibinfo{volume}{100}}, \bibinfo{pages}{063007}
  (\bibinfo{year}{2019}), \eprint{1904.09998}.

\bibitem[{\citenamefont{Kaplinghat et~al.}(2019)\citenamefont{Kaplinghat,
  Valli, and Yu}}]{Kaplinghat:2019svz}
\bibinfo{author}{\bibfnamefont{M.}~\bibnamefont{Kaplinghat}},
  \bibinfo{author}{\bibfnamefont{M.}~\bibnamefont{Valli}}, \bibnamefont{and}
  \bibinfo{author}{\bibfnamefont{H.-B.} \bibnamefont{Yu}},
  \bibinfo{journal}{Mon. Not. Roy. Astron. Soc.}
  \textbf{\bibinfo{volume}{490}}, \bibinfo{pages}{231} (\bibinfo{year}{2019}),
  \eprint{1904.04939}.

\bibitem[{\citenamefont{Kahlhoefer et~al.}(2019)\citenamefont{Kahlhoefer,
  Kaplinghat, Slatyer, and Wu}}]{Kahlhoefer:2019oyt}
\bibinfo{author}{\bibfnamefont{F.}~\bibnamefont{Kahlhoefer}},
  \bibinfo{author}{\bibfnamefont{M.}~\bibnamefont{Kaplinghat}},
  \bibinfo{author}{\bibfnamefont{T.~R.} \bibnamefont{Slatyer}},
  \bibnamefont{and} \bibinfo{author}{\bibfnamefont{C.-L.} \bibnamefont{Wu}},
  \bibinfo{journal}{JCAP} \textbf{\bibinfo{volume}{12}}, \bibinfo{pages}{010}
  (\bibinfo{year}{2019}), \eprint{1904.10539}.

\bibitem[{\citenamefont{Nadler et~al.}(2020)\citenamefont{Nadler, Banerjee,
  Adhikari, Mao, and Wechsler}}]{Nadler:2020ulu}
\bibinfo{author}{\bibfnamefont{E.~O.} \bibnamefont{Nadler}},
  \bibinfo{author}{\bibfnamefont{A.}~\bibnamefont{Banerjee}},
  \bibinfo{author}{\bibfnamefont{S.}~\bibnamefont{Adhikari}},
  \bibinfo{author}{\bibfnamefont{Y.-Y.} \bibnamefont{Mao}}, \bibnamefont{and}
  \bibinfo{author}{\bibfnamefont{R.~H.} \bibnamefont{Wechsler}},
  \bibinfo{journal}{Astrophys. J.} \textbf{\bibinfo{volume}{896}},
  \bibinfo{pages}{112} (\bibinfo{year}{2020}), \eprint{2001.08754}.

\bibitem[{\citenamefont{Turner et~al.}(2020)\citenamefont{Turner, Lovell,
  Zavala, and Vogelsberger}}]{Turner:2020vlf}
\bibinfo{author}{\bibfnamefont{H.~C.} \bibnamefont{Turner}},
  \bibinfo{author}{\bibfnamefont{M.~R.} \bibnamefont{Lovell}},
  \bibinfo{author}{\bibfnamefont{J.}~\bibnamefont{Zavala}}, \bibnamefont{and}
  \bibinfo{author}{\bibfnamefont{M.}~\bibnamefont{Vogelsberger}}
  (\bibinfo{year}{2020}), \eprint{2010.02924}.

\bibitem[{\citenamefont{{Correa}}(2021)}]{2021MNRAS.503..920C}
\bibinfo{author}{\bibfnamefont{C.~A.} \bibnamefont{{Correa}}},
  \bibinfo{journal}{Mon.~Not.~Roy.~Astron.~Soc.}
  \textbf{\bibinfo{volume}{503}}, \bibinfo{pages}{920} (\bibinfo{year}{2021}),
  \eprint{2007.02958}.

\bibitem[{\citenamefont{Sameie et~al.}(2021)\citenamefont{Sameie,
  Boylan-Kolchin, Sanderson, Vargya, Hopkins, Wetzel, Bullock, Graus, and
  Robles}}]{Sameie:2021ang}
\bibinfo{author}{\bibfnamefont{O.}~\bibnamefont{Sameie}},
  \bibinfo{author}{\bibfnamefont{M.}~\bibnamefont{Boylan-Kolchin}},
  \bibinfo{author}{\bibfnamefont{R.}~\bibnamefont{Sanderson}},
  \bibinfo{author}{\bibfnamefont{D.}~\bibnamefont{Vargya}},
  \bibinfo{author}{\bibfnamefont{P.}~\bibnamefont{Hopkins}},
  \bibinfo{author}{\bibfnamefont{A.}~\bibnamefont{Wetzel}},
  \bibinfo{author}{\bibfnamefont{J.}~\bibnamefont{Bullock}},
  \bibinfo{author}{\bibfnamefont{A.}~\bibnamefont{Graus}}, \bibnamefont{and}
  \bibinfo{author}{\bibfnamefont{V.}~\bibnamefont{Robles}}
  (\bibinfo{year}{2021}), \eprint{2102.12480}.

\bibitem[{\citenamefont{Sameie et~al.}(2020{\natexlab{a}})\citenamefont{Sameie,
  Chakrabarti, Yu, Boylan-Kolchin, Vogelsberger, Zavala, and
  Hernquist}}]{Sameie:2020xdi}
\bibinfo{author}{\bibfnamefont{O.}~\bibnamefont{Sameie}},
  \bibinfo{author}{\bibfnamefont{S.}~\bibnamefont{Chakrabarti}},
  \bibinfo{author}{\bibfnamefont{H.-B.} \bibnamefont{Yu}},
  \bibinfo{author}{\bibfnamefont{M.}~\bibnamefont{Boylan-Kolchin}},
  \bibinfo{author}{\bibfnamefont{M.}~\bibnamefont{Vogelsberger}},
  \bibinfo{author}{\bibfnamefont{J.}~\bibnamefont{Zavala}}, \bibnamefont{and}
  \bibinfo{author}{\bibfnamefont{L.}~\bibnamefont{Hernquist}}
  (\bibinfo{year}{2020}{\natexlab{a}}), \eprint{2006.06681}.

\bibitem[{\citenamefont{Salucci}(2019)}]{Salucci:2018hqu}
\bibinfo{author}{\bibfnamefont{P.}~\bibnamefont{Salucci}},
  \bibinfo{journal}{Astron. Astrophys. Rev.} \textbf{\bibinfo{volume}{27}},
  \bibinfo{pages}{2} (\bibinfo{year}{2019}), \eprint{1811.08843}.

\bibitem[{\citenamefont{{Kravtsov} et~al.}(2004)\citenamefont{{Kravtsov},
  {Gnedin}, and {Klypin}}}]{2004ApJ...609..482K}
\bibinfo{author}{\bibfnamefont{A.~V.} \bibnamefont{{Kravtsov}}},
  \bibinfo{author}{\bibfnamefont{O.~Y.} \bibnamefont{{Gnedin}}},
  \bibnamefont{and} \bibinfo{author}{\bibfnamefont{A.~A.}
  \bibnamefont{{Klypin}}}, \bibinfo{journal}{Astroph. J.}
  \textbf{\bibinfo{volume}{609}}, \bibinfo{pages}{482} (\bibinfo{year}{2004}),
  \eprint{astro-ph/0401088}.

\bibitem[{\citenamefont{{D'Onghia} et~al.}(2010)\citenamefont{{D'Onghia},
  {Springel}, {Hernquist}, and {Keres}}}]{2010ApJ...709.1138D}
\bibinfo{author}{\bibfnamefont{E.}~\bibnamefont{{D'Onghia}}},
  \bibinfo{author}{\bibfnamefont{V.}~\bibnamefont{{Springel}}},
  \bibinfo{author}{\bibfnamefont{L.}~\bibnamefont{{Hernquist}}},
  \bibnamefont{and} \bibinfo{author}{\bibfnamefont{D.}~\bibnamefont{{Keres}}},
  \bibinfo{journal}{Astroph. J.} \textbf{\bibinfo{volume}{709}},
  \bibinfo{pages}{1138} (\bibinfo{year}{2010}), \eprint{0907.3482}.

\bibitem[{\citenamefont{{Pe{\~n}arrubia}
  et~al.}(2010)\citenamefont{{Pe{\~n}arrubia}, {Benson}, {Walker}, {Gilmore},
  {McConnachie}, and {Mayer}}}]{2010MNRAS.406.1290P}
\bibinfo{author}{\bibfnamefont{J.}~\bibnamefont{{Pe{\~n}arrubia}}},
  \bibinfo{author}{\bibfnamefont{A.~J.} \bibnamefont{{Benson}}},
  \bibinfo{author}{\bibfnamefont{M.~G.} \bibnamefont{{Walker}}},
  \bibinfo{author}{\bibfnamefont{G.}~\bibnamefont{{Gilmore}}},
  \bibinfo{author}{\bibfnamefont{A.~W.} \bibnamefont{{McConnachie}}},
  \bibnamefont{and} \bibinfo{author}{\bibfnamefont{L.}~\bibnamefont{{Mayer}}},
  \bibinfo{journal}{Mon. Not. Roy. Astron. Soc.}
  \textbf{\bibinfo{volume}{406}}, \bibinfo{pages}{1290} (\bibinfo{year}{2010}),
  \eprint{1002.3376}.

\bibitem[{\citenamefont{{Dooley} et~al.}(2017)\citenamefont{{Dooley}, {Peter},
  {Yang}, {Willman}, {Griffen}, and {Frebel}}}]{2017MNRAS.471.4894D}
\bibinfo{author}{\bibfnamefont{G.~A.} \bibnamefont{{Dooley}}},
  \bibinfo{author}{\bibfnamefont{A.~H.~G.} \bibnamefont{{Peter}}},
  \bibinfo{author}{\bibfnamefont{T.}~\bibnamefont{{Yang}}},
  \bibinfo{author}{\bibfnamefont{B.}~\bibnamefont{{Willman}}},
  \bibinfo{author}{\bibfnamefont{B.~F.} \bibnamefont{{Griffen}}},
  \bibnamefont{and} \bibinfo{author}{\bibfnamefont{A.}~\bibnamefont{{Frebel}}},
  \bibinfo{journal}{Mon. Not. Roy. Astron. Soc.}
  \textbf{\bibinfo{volume}{471}}, \bibinfo{pages}{4894} (\bibinfo{year}{2017}),
  \eprint{1610.00708}.

\bibitem[{\citenamefont{{Garrison-Kimmel}
  et~al.}(2017)\citenamefont{{Garrison-Kimmel}, {Wetzel}, {Bullock}, {Hopkins},
  {Boylan-Kolchin}, {Faucher-Gigu{\`e}re}, {Kere{\v{s}}}, {Quataert},
  {Sanderson}, {Graus} et~al.}}]{2017MNRAS.471.1709G}
\bibinfo{author}{\bibfnamefont{S.}~\bibnamefont{{Garrison-Kimmel}}},
  \bibinfo{author}{\bibfnamefont{A.}~\bibnamefont{{Wetzel}}},
  \bibinfo{author}{\bibfnamefont{J.~S.} \bibnamefont{{Bullock}}},
  \bibinfo{author}{\bibfnamefont{P.~F.} \bibnamefont{{Hopkins}}},
  \bibinfo{author}{\bibfnamefont{M.}~\bibnamefont{{Boylan-Kolchin}}},
  \bibinfo{author}{\bibfnamefont{C.-A.} \bibnamefont{{Faucher-Gigu{\`e}re}}},
  \bibinfo{author}{\bibfnamefont{D.}~\bibnamefont{{Kere{\v{s}}}}},
  \bibinfo{author}{\bibfnamefont{E.}~\bibnamefont{{Quataert}}},
  \bibinfo{author}{\bibfnamefont{R.~E.} \bibnamefont{{Sanderson}}},
  \bibinfo{author}{\bibfnamefont{A.~S.} \bibnamefont{{Graus}}},
  \bibnamefont{et~al.}, \bibinfo{journal}{Mon. Not. Roy. Astron. Soc.}
  \textbf{\bibinfo{volume}{471}}, \bibinfo{pages}{1709} (\bibinfo{year}{2017}),
  \eprint{1701.03792}.

\bibitem[{\citenamefont{{Samuel} et~al.}(2020)\citenamefont{{Samuel}, {Wetzel},
  {Tollerud}, {Garrison-Kimmel}, {Loebman}, {El-Badry}, {Hopkins},
  {Boylan-Kolchin}, {Faucher-Gigu{\`e}re}, {Bullock}
  et~al.}}]{2020MNRAS.491.1471S}
\bibinfo{author}{\bibfnamefont{J.}~\bibnamefont{{Samuel}}},
  \bibinfo{author}{\bibfnamefont{A.}~\bibnamefont{{Wetzel}}},
  \bibinfo{author}{\bibfnamefont{E.}~\bibnamefont{{Tollerud}}},
  \bibinfo{author}{\bibfnamefont{S.}~\bibnamefont{{Garrison-Kimmel}}},
  \bibinfo{author}{\bibfnamefont{S.}~\bibnamefont{{Loebman}}},
  \bibinfo{author}{\bibfnamefont{K.}~\bibnamefont{{El-Badry}}},
  \bibinfo{author}{\bibfnamefont{P.~F.} \bibnamefont{{Hopkins}}},
  \bibinfo{author}{\bibfnamefont{M.}~\bibnamefont{{Boylan-Kolchin}}},
  \bibinfo{author}{\bibfnamefont{C.-A.} \bibnamefont{{Faucher-Gigu{\`e}re}}},
  \bibinfo{author}{\bibfnamefont{J.~S.} \bibnamefont{{Bullock}}},
  \bibnamefont{et~al.}, \bibinfo{journal}{Mon. Not. Roy. Astron. Soc.}
  \textbf{\bibinfo{volume}{491}}, \bibinfo{pages}{1471} (\bibinfo{year}{2020}),
  \eprint{1904.11508}.

\bibitem[{\citenamefont{Tulin and Yu}(2018)}]{Tulin:2017ara}
\bibinfo{author}{\bibfnamefont{S.}~\bibnamefont{Tulin}} \bibnamefont{and}
  \bibinfo{author}{\bibfnamefont{H.-B.} \bibnamefont{Yu}},
  \bibinfo{journal}{Phys. Rept.} \textbf{\bibinfo{volume}{730}},
  \bibinfo{pages}{1} (\bibinfo{year}{2018}), \eprint{1705.02358}.

\bibitem[{\citenamefont{Kamada et~al.}(2017)\citenamefont{Kamada, Kaplinghat,
  Pace, and Yu}}]{Kamada:2016euw}
\bibinfo{author}{\bibfnamefont{A.}~\bibnamefont{Kamada}},
  \bibinfo{author}{\bibfnamefont{M.}~\bibnamefont{Kaplinghat}},
  \bibinfo{author}{\bibfnamefont{A.~B.} \bibnamefont{Pace}}, \bibnamefont{and}
  \bibinfo{author}{\bibfnamefont{H.-B.} \bibnamefont{Yu}},
  \bibinfo{journal}{Phys. Rev. Lett.} \textbf{\bibinfo{volume}{119}},
  \bibinfo{pages}{111102} (\bibinfo{year}{2017}), \eprint{1611.02716}.

\bibitem[{\citenamefont{{Creasey} et~al.}(2017)\citenamefont{{Creasey},
  {Sameie}, {Sales}, {Yu}, {Vogelsberger}, and {Zavala}}}]{2017MNRAS.468.2283C}
\bibinfo{author}{\bibfnamefont{P.}~\bibnamefont{{Creasey}}},
  \bibinfo{author}{\bibfnamefont{O.}~\bibnamefont{{Sameie}}},
  \bibinfo{author}{\bibfnamefont{L.~V.} \bibnamefont{{Sales}}},
  \bibinfo{author}{\bibfnamefont{H.-B.} \bibnamefont{{Yu}}},
  \bibinfo{author}{\bibfnamefont{M.}~\bibnamefont{{Vogelsberger}}},
  \bibnamefont{and} \bibinfo{author}{\bibfnamefont{J.}~\bibnamefont{{Zavala}}},
  \bibinfo{journal}{Mon. Not. Roy. Astron. Soc.}
  \textbf{\bibinfo{volume}{468}}, \bibinfo{pages}{2283} (\bibinfo{year}{2017}),
  \eprint{1612.03903}.

\bibitem[{\citenamefont{Ren et~al.}(2019)\citenamefont{Ren, Kwa, Kaplinghat,
  and Yu}}]{Ren:2018jpt}
\bibinfo{author}{\bibfnamefont{T.}~\bibnamefont{Ren}},
  \bibinfo{author}{\bibfnamefont{A.}~\bibnamefont{Kwa}},
  \bibinfo{author}{\bibfnamefont{M.}~\bibnamefont{Kaplinghat}},
  \bibnamefont{and} \bibinfo{author}{\bibfnamefont{H.-B.} \bibnamefont{Yu}},
  \bibinfo{journal}{Phys. Rev. X} \textbf{\bibinfo{volume}{9}},
  \bibinfo{pages}{031020} (\bibinfo{year}{2019}), \eprint{1808.05695}.

\bibitem[{\citenamefont{Kaplinghat et~al.}(2020)\citenamefont{Kaplinghat, Ren,
  and Yu}}]{Kaplinghat:2019dhn}
\bibinfo{author}{\bibfnamefont{M.}~\bibnamefont{Kaplinghat}},
  \bibinfo{author}{\bibfnamefont{T.}~\bibnamefont{Ren}}, \bibnamefont{and}
  \bibinfo{author}{\bibfnamefont{H.-B.} \bibnamefont{Yu}},
  \bibinfo{journal}{JCAP} \textbf{\bibinfo{volume}{06}}, \bibinfo{pages}{027}
  (\bibinfo{year}{2020}), \eprint{1911.00544}.

\bibitem[{\citenamefont{Kummer et~al.}(2018)\citenamefont{Kummer, Kahlhoefer,
  and Schmidt-Hoberg}}]{Kummer:2017bhr}
\bibinfo{author}{\bibfnamefont{J.}~\bibnamefont{Kummer}},
  \bibinfo{author}{\bibfnamefont{F.}~\bibnamefont{Kahlhoefer}},
  \bibnamefont{and}
  \bibinfo{author}{\bibfnamefont{K.}~\bibnamefont{Schmidt-Hoberg}},
  \bibinfo{journal}{Mon. Not. Roy. Astron. Soc.}
  \textbf{\bibinfo{volume}{474}}, \bibinfo{pages}{388} (\bibinfo{year}{2018}),
  \eprint{1706.04794}.

\bibitem[{\citenamefont{Sameie et~al.}(2020{\natexlab{b}})\citenamefont{Sameie,
  Yu, Sales, Vogelsberger, and Zavala}}]{Sameie:2019zfo}
\bibinfo{author}{\bibfnamefont{O.}~\bibnamefont{Sameie}},
  \bibinfo{author}{\bibfnamefont{H.-B.} \bibnamefont{Yu}},
  \bibinfo{author}{\bibfnamefont{L.~V.} \bibnamefont{Sales}},
  \bibinfo{author}{\bibfnamefont{M.}~\bibnamefont{Vogelsberger}},
  \bibnamefont{and} \bibinfo{author}{\bibfnamefont{J.}~\bibnamefont{Zavala}},
  \bibinfo{journal}{Phys. Rev. Lett.} \textbf{\bibinfo{volume}{124}},
  \bibinfo{pages}{141102} (\bibinfo{year}{2020}{\natexlab{b}}),
  \eprint{1904.07872}.

\bibitem[{\citenamefont{Balberg et~al.}(2002)\citenamefont{Balberg, Shapiro,
  and Inagaki}}]{Balberg:2002ue}
\bibinfo{author}{\bibfnamefont{S.}~\bibnamefont{Balberg}},
  \bibinfo{author}{\bibfnamefont{S.~L.} \bibnamefont{Shapiro}},
  \bibnamefont{and} \bibinfo{author}{\bibfnamefont{S.}~\bibnamefont{Inagaki}},
  \bibinfo{journal}{Astrophys. J.} \textbf{\bibinfo{volume}{568}},
  \bibinfo{pages}{475} (\bibinfo{year}{2002}), \eprint{astro-ph/0110561}.

\bibitem[{\citenamefont{{Koda} and {Shapiro}}(2011)}]{2011MNRAS.415.1125K}
\bibinfo{author}{\bibfnamefont{J.}~\bibnamefont{{Koda}}} \bibnamefont{and}
  \bibinfo{author}{\bibfnamefont{P.~R.} \bibnamefont{{Shapiro}}},
  \bibinfo{journal}{Mon.~Not.~Roy.~Astron.~Soc.}
  \textbf{\bibinfo{volume}{415}}, \bibinfo{pages}{1125} (\bibinfo{year}{2011}),
  \eprint{1101.3097}.

\bibitem[{\citenamefont{Elbert et~al.}(2015)\citenamefont{Elbert, Bullock,
  Garrison-Kimmel, Rocha, O\~norbe, and Peter}}]{Elbert:2014bma}
\bibinfo{author}{\bibfnamefont{O.~D.} \bibnamefont{Elbert}},
  \bibinfo{author}{\bibfnamefont{J.~S.} \bibnamefont{Bullock}},
  \bibinfo{author}{\bibfnamefont{S.}~\bibnamefont{Garrison-Kimmel}},
  \bibinfo{author}{\bibfnamefont{M.}~\bibnamefont{Rocha}},
  \bibinfo{author}{\bibfnamefont{J.}~\bibnamefont{O\~norbe}}, \bibnamefont{and}
  \bibinfo{author}{\bibfnamefont{A.~H.~G.} \bibnamefont{Peter}},
  \bibinfo{journal}{Mon. Not. Roy. Astron. Soc.}
  \textbf{\bibinfo{volume}{453}}, \bibinfo{pages}{29} (\bibinfo{year}{2015}),
  \eprint{1412.1477}.

\bibitem[{\citenamefont{Essig et~al.}(2019)\citenamefont{Essig, Mcdermott, Yu,
  and Zhong}}]{Essig:2018pzq}
\bibinfo{author}{\bibfnamefont{R.}~\bibnamefont{Essig}},
  \bibinfo{author}{\bibfnamefont{S.~D.} \bibnamefont{Mcdermott}},
  \bibinfo{author}{\bibfnamefont{H.-B.} \bibnamefont{Yu}}, \bibnamefont{and}
  \bibinfo{author}{\bibfnamefont{Y.-M.} \bibnamefont{Zhong}},
  \bibinfo{journal}{Phys. Rev. Lett.} \textbf{\bibinfo{volume}{123}},
  \bibinfo{pages}{121102} (\bibinfo{year}{2019}), \eprint{1809.01144}.

\bibitem[{\citenamefont{Nishikawa et~al.}(2020)\citenamefont{Nishikawa, Boddy,
  and Kaplinghat}}]{Nishikawa:2019lsc}
\bibinfo{author}{\bibfnamefont{H.}~\bibnamefont{Nishikawa}},
  \bibinfo{author}{\bibfnamefont{K.~K.} \bibnamefont{Boddy}}, \bibnamefont{and}
  \bibinfo{author}{\bibfnamefont{M.}~\bibnamefont{Kaplinghat}},
  \bibinfo{journal}{Phys. Rev. D} \textbf{\bibinfo{volume}{101}},
  \bibinfo{pages}{063009} (\bibinfo{year}{2020}), \eprint{1901.00499}.

\bibitem[{\citenamefont{{Kim} and {Peter}}(2021)}]{2021arXiv210609050K}
\bibinfo{author}{\bibfnamefont{S.~Y.} \bibnamefont{{Kim}}} \bibnamefont{and}
  \bibinfo{author}{\bibfnamefont{A.~H.~G.} \bibnamefont{{Peter}}}
  (\bibinfo{year}{2021}), \eprint{2106.09050}.

\bibitem[{\citenamefont{Penarrubia et~al.}(2010)\citenamefont{Penarrubia,
  Benson, Walker, Gilmore, McConnachie, and Mayer}}]{Penarrubia:2010jk}
\bibinfo{author}{\bibfnamefont{J.}~\bibnamefont{Penarrubia}},
  \bibinfo{author}{\bibfnamefont{A.~J.} \bibnamefont{Benson}},
  \bibinfo{author}{\bibfnamefont{M.~G.} \bibnamefont{Walker}},
  \bibinfo{author}{\bibfnamefont{G.}~\bibnamefont{Gilmore}},
  \bibinfo{author}{\bibfnamefont{A.}~\bibnamefont{McConnachie}},
  \bibnamefont{and} \bibinfo{author}{\bibfnamefont{L.}~\bibnamefont{Mayer}},
  \bibinfo{journal}{Mon. Not. Roy. Astron. Soc.}
  \textbf{\bibinfo{volume}{406}}, \bibinfo{pages}{1290} (\bibinfo{year}{2010}),
  \eprint{1002.3376}.

\bibitem[{\citenamefont{Feng et~al.}(2009)\citenamefont{Feng, Kaplinghat, Tu,
  and Yu}}]{Feng:2009mn}
\bibinfo{author}{\bibfnamefont{J.~L.} \bibnamefont{Feng}},
  \bibinfo{author}{\bibfnamefont{M.}~\bibnamefont{Kaplinghat}},
  \bibinfo{author}{\bibfnamefont{H.}~\bibnamefont{Tu}}, \bibnamefont{and}
  \bibinfo{author}{\bibfnamefont{H.-B.} \bibnamefont{Yu}},
  \bibinfo{journal}{JCAP} \textbf{\bibinfo{volume}{07}}, \bibinfo{pages}{004}
  (\bibinfo{year}{2009}), \eprint{0905.3039}.

\bibitem[{\citenamefont{Loeb and Weiner}(2011)}]{Loeb:2010gj}
\bibinfo{author}{\bibfnamefont{A.}~\bibnamefont{Loeb}} \bibnamefont{and}
  \bibinfo{author}{\bibfnamefont{N.}~\bibnamefont{Weiner}},
  \bibinfo{journal}{Phys. Rev. Lett.} \textbf{\bibinfo{volume}{106}},
  \bibinfo{pages}{171302} (\bibinfo{year}{2011}), \eprint{1011.6374}.

\bibitem[{\citenamefont{Kaplinghat et~al.}(2016)\citenamefont{Kaplinghat,
  Tulin, and Yu}}]{Kaplinghat:2015aga}
\bibinfo{author}{\bibfnamefont{M.}~\bibnamefont{Kaplinghat}},
  \bibinfo{author}{\bibfnamefont{S.}~\bibnamefont{Tulin}}, \bibnamefont{and}
  \bibinfo{author}{\bibfnamefont{H.-B.} \bibnamefont{Yu}},
  \bibinfo{journal}{Phys. Rev. Lett.} \textbf{\bibinfo{volume}{116}},
  \bibinfo{pages}{041302} (\bibinfo{year}{2016}), \eprint{1508.03339}.

\bibitem[{\citenamefont{Spergel and Steinhardt}(2000)}]{Spergel:1999mh}
\bibinfo{author}{\bibfnamefont{D.~N.} \bibnamefont{Spergel}} \bibnamefont{and}
  \bibinfo{author}{\bibfnamefont{P.~J.} \bibnamefont{Steinhardt}},
  \bibinfo{journal}{Phys. Rev. Lett.} \textbf{\bibinfo{volume}{84}},
  \bibinfo{pages}{3760} (\bibinfo{year}{2000}), \eprint{astro-ph/9909386}.

\bibitem[{\citenamefont{{Navarro} et~al.}(1997)\citenamefont{{Navarro},
  {Frenk}, and {White}}}]{1997ApJ...490..493N}
\bibinfo{author}{\bibfnamefont{J.~F.} \bibnamefont{{Navarro}}},
  \bibinfo{author}{\bibfnamefont{C.~S.} \bibnamefont{{Frenk}}},
  \bibnamefont{and} \bibinfo{author}{\bibfnamefont{S.~D.~M.}
  \bibnamefont{{White}}}, \bibinfo{journal}{Astrophys. J.}
  \textbf{\bibinfo{volume}{490}}, \bibinfo{pages}{493} (\bibinfo{year}{1997}),
  \eprint{astro-ph/9611107}.

\bibitem[{\citenamefont{{Kaplinghat} et~al.}(2014)\citenamefont{{Kaplinghat},
  {Keeley}, {Linden}, and {Yu}}}]{2014PhRvL.113b1302K}
\bibinfo{author}{\bibfnamefont{M.}~\bibnamefont{{Kaplinghat}}},
  \bibinfo{author}{\bibfnamefont{R.~E.} \bibnamefont{{Keeley}}},
  \bibinfo{author}{\bibfnamefont{T.}~\bibnamefont{{Linden}}}, \bibnamefont{and}
  \bibinfo{author}{\bibfnamefont{H.-B.} \bibnamefont{{Yu}}},
  \bibinfo{journal}{\prl} \textbf{\bibinfo{volume}{113}}, \bibinfo{eid}{021302}
  (\bibinfo{year}{2014}), \eprint{1311.6524}.

\bibitem[{\citenamefont{Sameie et~al.}(2018)\citenamefont{Sameie, Creasey, Yu,
  Sales, Vogelsberger, and Zavala}}]{Sameie:2018chj}
\bibinfo{author}{\bibfnamefont{O.}~\bibnamefont{Sameie}},
  \bibinfo{author}{\bibfnamefont{P.}~\bibnamefont{Creasey}},
  \bibinfo{author}{\bibfnamefont{H.-B.} \bibnamefont{Yu}},
  \bibinfo{author}{\bibfnamefont{L.~V.} \bibnamefont{Sales}},
  \bibinfo{author}{\bibfnamefont{M.}~\bibnamefont{Vogelsberger}},
  \bibnamefont{and} \bibinfo{author}{\bibfnamefont{J.}~\bibnamefont{Zavala}},
  \bibinfo{journal}{Mon. Not. Roy. Astron. Soc.}
  \textbf{\bibinfo{volume}{479}}, \bibinfo{pages}{359} (\bibinfo{year}{2018}),
  \eprint{1801.09682}.

\bibitem[{\citenamefont{{Kelley} et~al.}(2019)\citenamefont{{Kelley},
  {Bullock}, {Garrison-Kimmel}, {Boylan-Kolchin}, {Pawlowski}, and
  {Graus}}}]{2019MNRAS.487.4409K}
\bibinfo{author}{\bibfnamefont{T.}~\bibnamefont{{Kelley}}},
  \bibinfo{author}{\bibfnamefont{J.~S.} \bibnamefont{{Bullock}}},
  \bibinfo{author}{\bibfnamefont{S.}~\bibnamefont{{Garrison-Kimmel}}},
  \bibinfo{author}{\bibfnamefont{M.}~\bibnamefont{{Boylan-Kolchin}}},
  \bibinfo{author}{\bibfnamefont{M.~S.} \bibnamefont{{Pawlowski}}},
  \bibnamefont{and} \bibinfo{author}{\bibfnamefont{A.~S.}
  \bibnamefont{{Graus}}}, \bibinfo{journal}{Mon. Not. Roy. Astron. Soc.}
  \textbf{\bibinfo{volume}{487}}, \bibinfo{pages}{4409} (\bibinfo{year}{2019}),
  \eprint{1811.12413}.

\bibitem[{\citenamefont{{Green} et~al.}(2021)\citenamefont{{Green}, {van den
  Bosch}, and {Jiang}}}]{2021MNRAS.503.4075G}
\bibinfo{author}{\bibfnamefont{S.~B.} \bibnamefont{{Green}}},
  \bibinfo{author}{\bibfnamefont{F.~C.} \bibnamefont{{van den Bosch}}},
  \bibnamefont{and} \bibinfo{author}{\bibfnamefont{F.}~\bibnamefont{{Jiang}}},
  \bibinfo{journal}{MNRAS} \textbf{\bibinfo{volume}{503}},
  \bibinfo{pages}{4075} (\bibinfo{year}{2021}), \eprint{2103.01227}.

\bibitem[{\citenamefont{{von Hoerner}}(1957)}]{1957ApJ...125..451V}
\bibinfo{author}{\bibfnamefont{S.}~\bibnamefont{{von Hoerner}}},
  \bibinfo{journal}{Astrophys. J.} \textbf{\bibinfo{volume}{125}},
  \bibinfo{pages}{451} (\bibinfo{year}{1957}).

\bibitem[{\citenamefont{{King}}(1962)}]{1962AJ.....67..471K}
\bibinfo{author}{\bibfnamefont{I.}~\bibnamefont{{King}}},
  \bibinfo{journal}{Astron. J.} \textbf{\bibinfo{volume}{67}},
  \bibinfo{pages}{471} (\bibinfo{year}{1962}).

\bibitem[{\citenamefont{Green et~al.}(2021)\citenamefont{Green, van~den Bosch,
  and Jiang}}]{10.1093/mnras/stab696}
\bibinfo{author}{\bibfnamefont{S.~B.} \bibnamefont{Green}},
  \bibinfo{author}{\bibfnamefont{F.~C.} \bibnamefont{van~den Bosch}},
  \bibnamefont{and} \bibinfo{author}{\bibfnamefont{F.}~\bibnamefont{Jiang}},
  \bibinfo{journal}{Mon.~Not.~Roy.~Astron.~Soc.}
  \textbf{\bibinfo{volume}{503}}, \bibinfo{pages}{4075} (\bibinfo{year}{2021}),
  ISSN \bibinfo{issn}{0035-8711}, \eprint{2103.01227}.

\bibitem[{\citenamefont{Jiang et~al.}(2021)\citenamefont{Jiang, Dekel,
  Freundlich, van~den Bosch, Green, Hopkins, Benson, and Du}}]{Jiang:2020rdj}
\bibinfo{author}{\bibfnamefont{F.}~\bibnamefont{Jiang}},
  \bibinfo{author}{\bibfnamefont{A.}~\bibnamefont{Dekel}},
  \bibinfo{author}{\bibfnamefont{J.}~\bibnamefont{Freundlich}},
  \bibinfo{author}{\bibfnamefont{F.~C.} \bibnamefont{van~den Bosch}},
  \bibinfo{author}{\bibfnamefont{S.~B.} \bibnamefont{Green}},
  \bibinfo{author}{\bibfnamefont{P.~F.} \bibnamefont{Hopkins}},
  \bibinfo{author}{\bibfnamefont{A.}~\bibnamefont{Benson}}, \bibnamefont{and}
  \bibinfo{author}{\bibfnamefont{X.}~\bibnamefont{Du}}, \bibinfo{journal}{Mon.
  Not. Roy. Astron. Soc.} \textbf{\bibinfo{volume}{502}}, \bibinfo{pages}{621}
  (\bibinfo{year}{2021}), \eprint{2005.05974}.

\bibitem[{\citenamefont{{Chandrasekhar}}(1943)}]{1943ApJ....97..255C}
\bibinfo{author}{\bibfnamefont{S.}~\bibnamefont{{Chandrasekhar}}},
  \bibinfo{journal}{Astrophys. J.} \textbf{\bibinfo{volume}{97}},
  \bibinfo{pages}{255} (\bibinfo{year}{1943}).

\bibitem[{\citenamefont{{Dutton} and {Macci{\`o}}}(2014)}]{2014MNRAS.441.3359D}
\bibinfo{author}{\bibfnamefont{A.~A.} \bibnamefont{{Dutton}}} \bibnamefont{and}
  \bibinfo{author}{\bibfnamefont{A.~V.} \bibnamefont{{Macci{\`o}}}},
  \bibinfo{journal}{Mon. Not. Roy. Astron. Soc.}
  \textbf{\bibinfo{volume}{441}}, \bibinfo{pages}{3359} (\bibinfo{year}{2014}),
  \eprint{1402.7073}.

\bibitem[{\citenamefont{{Rocha} et~al.}(2013)\citenamefont{{Rocha}, {Peter},
  {Bullock}, {Kaplinghat}, {Garrison-Kimmel}, {O{\~n}orbe}, and
  {Moustakas}}}]{2013MNRAS.430...81R}
\bibinfo{author}{\bibfnamefont{M.}~\bibnamefont{{Rocha}}},
  \bibinfo{author}{\bibfnamefont{A.~H.~G.} \bibnamefont{{Peter}}},
  \bibinfo{author}{\bibfnamefont{J.~S.} \bibnamefont{{Bullock}}},
  \bibinfo{author}{\bibfnamefont{M.}~\bibnamefont{{Kaplinghat}}},
  \bibinfo{author}{\bibfnamefont{S.}~\bibnamefont{{Garrison-Kimmel}}},
  \bibinfo{author}{\bibfnamefont{J.}~\bibnamefont{{O{\~n}orbe}}},
  \bibnamefont{and} \bibinfo{author}{\bibfnamefont{L.~A.}
  \bibnamefont{{Moustakas}}}, \bibinfo{journal}{MNRAS}
  \textbf{\bibinfo{volume}{430}}, \bibinfo{pages}{81} (\bibinfo{year}{2013}),
  \eprint{1208.3025}.

\bibitem[{\citenamefont{Sagunski et~al.}(2021)\citenamefont{Sagunski, Gad-Nasr,
  Colquhoun, Robertson, and Tulin}}]{Sagunski:2020spe}
\bibinfo{author}{\bibfnamefont{L.}~\bibnamefont{Sagunski}},
  \bibinfo{author}{\bibfnamefont{S.}~\bibnamefont{Gad-Nasr}},
  \bibinfo{author}{\bibfnamefont{B.}~\bibnamefont{Colquhoun}},
  \bibinfo{author}{\bibfnamefont{A.}~\bibnamefont{Robertson}},
  \bibnamefont{and} \bibinfo{author}{\bibfnamefont{S.}~\bibnamefont{Tulin}},
  \bibinfo{journal}{JCAP} \textbf{\bibinfo{volume}{01}}, \bibinfo{pages}{024}
  (\bibinfo{year}{2021}), \eprint{2006.12515}.

\bibitem[{\citenamefont{Patel et~al.}(2020)\citenamefont{Patel, Kallivayalil,
  Garavito-Camargo, Besla, Weisz, van~der Marel, Boylan-Kolchin, Pawlowski, and
  G{\'{o}}mez}}]{Patel_2020}
\bibinfo{author}{\bibfnamefont{E.}~\bibnamefont{Patel}},
  \bibinfo{author}{\bibfnamefont{N.}~\bibnamefont{Kallivayalil}},
  \bibinfo{author}{\bibfnamefont{N.}~\bibnamefont{Garavito-Camargo}},
  \bibinfo{author}{\bibfnamefont{G.}~\bibnamefont{Besla}},
  \bibinfo{author}{\bibfnamefont{D.~R.} \bibnamefont{Weisz}},
  \bibinfo{author}{\bibfnamefont{R.~P.} \bibnamefont{van~der Marel}},
  \bibinfo{author}{\bibfnamefont{M.}~\bibnamefont{Boylan-Kolchin}},
  \bibinfo{author}{\bibfnamefont{M.~S.} \bibnamefont{Pawlowski}},
  \bibnamefont{and} \bibinfo{author}{\bibfnamefont{F.~A.}
  \bibnamefont{G{\'{o}}mez}}, \bibinfo{journal}{Astrophys. J.}
  \textbf{\bibinfo{volume}{893}}, \bibinfo{pages}{121} (\bibinfo{year}{2020}).

\bibitem[{\citenamefont{Bland-Hawthorn and
  Gerhard}(2016)}]{Bland_Hawthorn_2016}
\bibinfo{author}{\bibfnamefont{J.}~\bibnamefont{Bland-Hawthorn}}
  \bibnamefont{and} \bibinfo{author}{\bibfnamefont{O.}~\bibnamefont{Gerhard}},
  \bibinfo{journal}{Annual Review of Astronomy and Astrophysics}
  \textbf{\bibinfo{volume}{54}}, \bibinfo{pages}{529} (\bibinfo{year}{2016}).

\bibitem[{\citenamefont{Harvey et~al.}(2019)\citenamefont{Harvey, Robertson,
  Massey, and McCarthy}}]{Harvey:2018uwf}
\bibinfo{author}{\bibfnamefont{D.}~\bibnamefont{Harvey}},
  \bibinfo{author}{\bibfnamefont{A.}~\bibnamefont{Robertson}},
  \bibinfo{author}{\bibfnamefont{R.}~\bibnamefont{Massey}}, \bibnamefont{and}
  \bibinfo{author}{\bibfnamefont{I.~G.} \bibnamefont{McCarthy}},
  \bibinfo{journal}{Mon. Not. Roy. Astron. Soc.}
  \textbf{\bibinfo{volume}{488}}, \bibinfo{pages}{1572} (\bibinfo{year}{2019}),
  \eprint{1812.06981}.

\bibitem[{\citenamefont{Andrade et~al.}(2020)\citenamefont{Andrade, Fuson,
  Gad-Nasr, Kong, Minor, Roberts, and Kaplinghat}}]{Andrade:2020lqq}
\bibinfo{author}{\bibfnamefont{K.~E.} \bibnamefont{Andrade}},
  \bibinfo{author}{\bibfnamefont{J.}~\bibnamefont{Fuson}},
  \bibinfo{author}{\bibfnamefont{S.}~\bibnamefont{Gad-Nasr}},
  \bibinfo{author}{\bibfnamefont{D.}~\bibnamefont{Kong}},
  \bibinfo{author}{\bibfnamefont{Q.}~\bibnamefont{Minor}},
  \bibinfo{author}{\bibfnamefont{M.~G.} \bibnamefont{Roberts}},
  \bibnamefont{and}
  \bibinfo{author}{\bibfnamefont{M.}~\bibnamefont{Kaplinghat}}
  (\bibinfo{year}{2020}), \eprint{2012.06611}.

\bibitem[{\citenamefont{Meneghetti et~al.}(2020)}]{Meneghetti:2020yif}
\bibinfo{author}{\bibfnamefont{M.}~\bibnamefont{Meneghetti}}
  \bibnamefont{et~al.}, \bibinfo{journal}{Science}
  \textbf{\bibinfo{volume}{369}}, \bibinfo{pages}{1347} (\bibinfo{year}{2020}),
  \eprint{2009.04471}.

\bibitem[{\citenamefont{Minor et~al.}(2020)\citenamefont{Minor, Gad-Nasr,
  Kaplinghat, and Vegetti}}]{Minor:2020hic}
\bibinfo{author}{\bibfnamefont{Q.~E.} \bibnamefont{Minor}},
  \bibinfo{author}{\bibfnamefont{S.}~\bibnamefont{Gad-Nasr}},
  \bibinfo{author}{\bibfnamefont{M.}~\bibnamefont{Kaplinghat}},
  \bibnamefont{and} \bibinfo{author}{\bibfnamefont{S.}~\bibnamefont{Vegetti}}
  (\bibinfo{year}{2020}), \eprint{2011.10627}.

\bibitem[{\citenamefont{Yang and Yu}(2021)}]{Yang:2021kdf}
\bibinfo{author}{\bibfnamefont{D.}~\bibnamefont{Yang}} \bibnamefont{and}
  \bibinfo{author}{\bibfnamefont{H.-B.} \bibnamefont{Yu}}
  (\bibinfo{year}{2021}), \eprint{2102.02375}.

\bibitem[{\citenamefont{{Pe{\~n}arrubia}
  et~al.}(2008)\citenamefont{{Pe{\~n}arrubia}, {Navarro}, and
  {McConnachie}}}]{2008AN....329..934P}
\bibinfo{author}{\bibfnamefont{J.}~\bibnamefont{{Pe{\~n}arrubia}}},
  \bibinfo{author}{\bibfnamefont{J.~F.} \bibnamefont{{Navarro}}},
  \bibnamefont{and} \bibinfo{author}{\bibfnamefont{A.~W.}
  \bibnamefont{{McConnachie}}}, \bibinfo{journal}{Astronomische Nachrichten}
  \textbf{\bibinfo{volume}{329}}, \bibinfo{pages}{934} (\bibinfo{year}{2008}).

\bibitem[{\citenamefont{{Hayashi} et~al.}(2003)\citenamefont{{Hayashi},
  {Navarro}, {Taylor}, {Stadel}, and {Quinn}}}]{2003ApJ...584..541H}
\bibinfo{author}{\bibfnamefont{E.}~\bibnamefont{{Hayashi}}},
  \bibinfo{author}{\bibfnamefont{J.~F.} \bibnamefont{{Navarro}}},
  \bibinfo{author}{\bibfnamefont{J.~E.} \bibnamefont{{Taylor}}},
  \bibinfo{author}{\bibfnamefont{J.}~\bibnamefont{{Stadel}}}, \bibnamefont{and}
  \bibinfo{author}{\bibfnamefont{T.}~\bibnamefont{{Quinn}}},
  \bibinfo{journal}{\apj} \textbf{\bibinfo{volume}{584}}, \bibinfo{pages}{541}
  (\bibinfo{year}{2003}), \eprint{astro-ph/0203004}.

\bibitem[{\citenamefont{Helmi}(2020)}]{2020ARA_A..58..205H}
\bibinfo{author}{\bibfnamefont{A.}~\bibnamefont{Helmi}},
  \bibinfo{journal}{araa} \textbf{\bibinfo{volume}{58}}, \bibinfo{pages}{205}
  (\bibinfo{year}{2020}), \eprint{2002.04340}.

\bibitem[{\citenamefont{Neto et~al.}(2007)\citenamefont{Neto, Gao, Bett, Cole,
  Navarro, Frenk, White, Springel, and Jenkins}}]{Neto:2007vq}
\bibinfo{author}{\bibfnamefont{A.~F.} \bibnamefont{Neto}},
  \bibinfo{author}{\bibfnamefont{L.}~\bibnamefont{Gao}},
  \bibinfo{author}{\bibfnamefont{P.}~\bibnamefont{Bett}},
  \bibinfo{author}{\bibfnamefont{S.}~\bibnamefont{Cole}},
  \bibinfo{author}{\bibfnamefont{J.~F.} \bibnamefont{Navarro}},
  \bibinfo{author}{\bibfnamefont{C.~S.} \bibnamefont{Frenk}},
  \bibinfo{author}{\bibfnamefont{S.~D.~M.} \bibnamefont{White}},
  \bibinfo{author}{\bibfnamefont{V.}~\bibnamefont{Springel}}, \bibnamefont{and}
  \bibinfo{author}{\bibfnamefont{A.}~\bibnamefont{Jenkins}},
  \bibinfo{journal}{Mon. Not. Roy. Astron. Soc.}
  \textbf{\bibinfo{volume}{381}}, \bibinfo{pages}{1450} (\bibinfo{year}{2007}),
  \eprint{0706.2919}.

\bibitem[{\citenamefont{Diemer and Joyce}(2019)}]{Diemer:2018vmz}
\bibinfo{author}{\bibfnamefont{B.}~\bibnamefont{Diemer}} \bibnamefont{and}
  \bibinfo{author}{\bibfnamefont{M.}~\bibnamefont{Joyce}},
  \bibinfo{journal}{Astrophys. J.} \textbf{\bibinfo{volume}{871}},
  \bibinfo{pages}{168} (\bibinfo{year}{2019}), \eprint{1809.07326}.

\bibitem[{\citenamefont{Li et~al.}(2020)\citenamefont{Li, Qian, Han, Li, Wang,
  and Jing}}]{Li_2020}
\bibinfo{author}{\bibfnamefont{Z.-Z.} \bibnamefont{Li}},
  \bibinfo{author}{\bibfnamefont{Y.-Z.} \bibnamefont{Qian}},
  \bibinfo{author}{\bibfnamefont{J.}~\bibnamefont{Han}},
  \bibinfo{author}{\bibfnamefont{T.~S.} \bibnamefont{Li}},
  \bibinfo{author}{\bibfnamefont{W.}~\bibnamefont{Wang}}, \bibnamefont{and}
  \bibinfo{author}{\bibfnamefont{Y.~P.} \bibnamefont{Jing}},
  \bibinfo{journal}{Astrophys. J.} \textbf{\bibinfo{volume}{894}},
  \bibinfo{pages}{10} (\bibinfo{year}{2020}).

\bibitem[{\citenamefont{Tulin et~al.}(2013)\citenamefont{Tulin, Yu, and
  Zurek}}]{Tulin:2013teo}
\bibinfo{author}{\bibfnamefont{S.}~\bibnamefont{Tulin}},
  \bibinfo{author}{\bibfnamefont{H.-B.} \bibnamefont{Yu}}, \bibnamefont{and}
  \bibinfo{author}{\bibfnamefont{K.~M.} \bibnamefont{Zurek}},
  \bibinfo{journal}{Phys. Rev. D} \textbf{\bibinfo{volume}{87}},
  \bibinfo{pages}{115007} (\bibinfo{year}{2013}), \eprint{1302.3898}.

\bibitem[{\citenamefont{Robertson et~al.}(2017)\citenamefont{Robertson, Massey,
  and Eke}}]{Robertson:2016qef}
\bibinfo{author}{\bibfnamefont{A.}~\bibnamefont{Robertson}},
  \bibinfo{author}{\bibfnamefont{R.}~\bibnamefont{Massey}}, \bibnamefont{and}
  \bibinfo{author}{\bibfnamefont{V.}~\bibnamefont{Eke}}, \bibinfo{journal}{Mon.
  Not. Roy. Astron. Soc.} \textbf{\bibinfo{volume}{467}}, \bibinfo{pages}{4719}
  (\bibinfo{year}{2017}), \eprint{1612.03906}.

\bibitem[{\citenamefont{Robertson et~al.}(2019)\citenamefont{Robertson, Harvey,
  Massey, Eke, McCarthy, Jauzac, Li, and Schaye}}]{Robertson:2018anx}
\bibinfo{author}{\bibfnamefont{A.}~\bibnamefont{Robertson}},
  \bibinfo{author}{\bibfnamefont{D.}~\bibnamefont{Harvey}},
  \bibinfo{author}{\bibfnamefont{R.}~\bibnamefont{Massey}},
  \bibinfo{author}{\bibfnamefont{V.}~\bibnamefont{Eke}},
  \bibinfo{author}{\bibfnamefont{I.~G.} \bibnamefont{McCarthy}},
  \bibinfo{author}{\bibfnamefont{M.}~\bibnamefont{Jauzac}},
  \bibinfo{author}{\bibfnamefont{B.}~\bibnamefont{Li}}, \bibnamefont{and}
  \bibinfo{author}{\bibfnamefont{J.}~\bibnamefont{Schaye}},
  \bibinfo{journal}{Mon. Not. Roy. Astron. Soc.}
  \textbf{\bibinfo{volume}{488}}, \bibinfo{pages}{3646} (\bibinfo{year}{2019}),
  \eprint{1810.05649}.

\bibitem[{\citenamefont{{Gan} et~al.}(2010)\citenamefont{{Gan}, {Kang}, {van
  den Bosch}, and {Hou}}}]{2010MNRAS.408.2201G}
\bibinfo{author}{\bibfnamefont{J.}~\bibnamefont{{Gan}}},
  \bibinfo{author}{\bibfnamefont{X.}~\bibnamefont{{Kang}}},
  \bibinfo{author}{\bibfnamefont{F.~C.} \bibnamefont{{van den Bosch}}},
  \bibnamefont{and} \bibinfo{author}{\bibfnamefont{J.}~\bibnamefont{{Hou}}},
  \bibinfo{journal}{MNRAS} \textbf{\bibinfo{volume}{408}},
  \bibinfo{pages}{2201} (\bibinfo{year}{2010}), \eprint{1007.0023}.

\bibitem[{\citenamefont{Martinez et~al.}(2011)\citenamefont{Martinez, Minor,
  Bullock, Kaplinghat, Simon, and Geha}}]{Martinez:2010xn}
\bibinfo{author}{\bibfnamefont{G.~D.} \bibnamefont{Martinez}},
  \bibinfo{author}{\bibfnamefont{Q.~E.} \bibnamefont{Minor}},
  \bibinfo{author}{\bibfnamefont{J.}~\bibnamefont{Bullock}},
  \bibinfo{author}{\bibfnamefont{M.}~\bibnamefont{Kaplinghat}},
  \bibinfo{author}{\bibfnamefont{J.~D.} \bibnamefont{Simon}}, \bibnamefont{and}
  \bibinfo{author}{\bibfnamefont{M.}~\bibnamefont{Geha}},
  \bibinfo{journal}{Astrophys. J.} \textbf{\bibinfo{volume}{738}},
  \bibinfo{pages}{55} (\bibinfo{year}{2011}), \eprint{1008.4585}.

\bibitem[{\citenamefont{Walker et~al.}(2016)\citenamefont{Walker, Mateo,
  Olszewski, Koposov, Belokurov, Jethwa, Nidever, Bonnivard, III, Bell
  et~al.}}]{Walker_2016}
\bibinfo{author}{\bibfnamefont{M.~G.} \bibnamefont{Walker}},
  \bibinfo{author}{\bibfnamefont{M.}~\bibnamefont{Mateo}},
  \bibinfo{author}{\bibfnamefont{E.~W.} \bibnamefont{Olszewski}},
  \bibinfo{author}{\bibfnamefont{S.}~\bibnamefont{Koposov}},
  \bibinfo{author}{\bibfnamefont{V.}~\bibnamefont{Belokurov}},
  \bibinfo{author}{\bibfnamefont{P.}~\bibnamefont{Jethwa}},
  \bibinfo{author}{\bibfnamefont{D.~L.} \bibnamefont{Nidever}},
  \bibinfo{author}{\bibfnamefont{V.}~\bibnamefont{Bonnivard}},
  \bibinfo{author}{\bibfnamefont{J.~I.~B.} \bibnamefont{III}},
  \bibinfo{author}{\bibfnamefont{E.~F.} \bibnamefont{Bell}},
  \bibnamefont{et~al.}, \bibinfo{journal}{Astrophys. J.}
  \textbf{\bibinfo{volume}{819}}, \bibinfo{pages}{53} (\bibinfo{year}{2016}).

\bibitem[{\citenamefont{Read and Erkal}(2019)}]{Read:2018gpi}
\bibinfo{author}{\bibfnamefont{J.~I.} \bibnamefont{Read}} \bibnamefont{and}
  \bibinfo{author}{\bibfnamefont{D.}~\bibnamefont{Erkal}},
  \bibinfo{journal}{Mon. Not. Roy. Astron. Soc.}
  \textbf{\bibinfo{volume}{487}}, \bibinfo{pages}{5799} (\bibinfo{year}{2019}),
  \eprint{1807.07093}.

\bibitem[{\citenamefont{Grand et~al.}(2021)\citenamefont{Grand, Marinacci,
  Pakmor, Simpson, Kelly, G\'omez, Jenkins, Springel, Frenk, and
  White}}]{Grand:2021fpx}
\bibinfo{author}{\bibfnamefont{R.~J.~J.} \bibnamefont{Grand}},
  \bibinfo{author}{\bibfnamefont{F.}~\bibnamefont{Marinacci}},
  \bibinfo{author}{\bibfnamefont{R.}~\bibnamefont{Pakmor}},
  \bibinfo{author}{\bibfnamefont{C.~M.} \bibnamefont{Simpson}},
  \bibinfo{author}{\bibfnamefont{A.~J.} \bibnamefont{Kelly}},
  \bibinfo{author}{\bibfnamefont{F.~A.} \bibnamefont{G\'omez}},
  \bibinfo{author}{\bibfnamefont{A.}~\bibnamefont{Jenkins}},
  \bibinfo{author}{\bibfnamefont{V.}~\bibnamefont{Springel}},
  \bibinfo{author}{\bibfnamefont{C.~S.} \bibnamefont{Frenk}}, \bibnamefont{and}
  \bibinfo{author}{\bibfnamefont{S.~D.~M.} \bibnamefont{White}}
  (\bibinfo{year}{2021}), \eprint{2105.04560}.

\bibitem[{\citenamefont{{Fillingham} et~al.}(2019)\citenamefont{{Fillingham},
  {Cooper}, {Kelley}, {Rodriguez Wimberly}, {Boylan-Kolchin}, {Bullock},
  {Garrison-Kimmel}, {Pawlowski}, and {Wheeler}}}]{2019arXiv190604180F}
\bibinfo{author}{\bibfnamefont{S.~P.} \bibnamefont{{Fillingham}}},
  \bibinfo{author}{\bibfnamefont{M.~C.} \bibnamefont{{Cooper}}},
  \bibinfo{author}{\bibfnamefont{T.}~\bibnamefont{{Kelley}}},
  \bibinfo{author}{\bibfnamefont{M.~K.} \bibnamefont{{Rodriguez Wimberly}}},
  \bibinfo{author}{\bibfnamefont{M.}~\bibnamefont{{Boylan-Kolchin}}},
  \bibinfo{author}{\bibfnamefont{J.~S.} \bibnamefont{{Bullock}}},
  \bibinfo{author}{\bibfnamefont{S.}~\bibnamefont{{Garrison-Kimmel}}},
  \bibinfo{author}{\bibfnamefont{M.~S.} \bibnamefont{{Pawlowski}}},
  \bibnamefont{and} \bibinfo{author}{\bibfnamefont{C.}~\bibnamefont{{Wheeler}}}
  (\bibinfo{year}{2019}), \eprint{1906.04180}.

\bibitem[{\citenamefont{{D'Souza} and {Bell}}(2021)}]{2021MNRAS.504.5270D}
\bibinfo{author}{\bibfnamefont{R.}~\bibnamefont{{D'Souza}}} \bibnamefont{and}
  \bibinfo{author}{\bibfnamefont{E.~F.} \bibnamefont{{Bell}}},
  \bibinfo{journal}{Mon.~Not.~Roy.~Astron.~Soc.}
  \textbf{\bibinfo{volume}{504}}, \bibinfo{pages}{5270} (\bibinfo{year}{2021}),
  \eprint{2104.13249}.

\bibitem[{\citenamefont{Ludlow et~al.}(2016)\citenamefont{Ludlow, Bose, Angulo,
  Wang, Hellwing, Navarro, Cole, and Frenk}}]{Ludlow:2016ifl}
\bibinfo{author}{\bibfnamefont{A.~D.} \bibnamefont{Ludlow}},
  \bibinfo{author}{\bibfnamefont{S.}~\bibnamefont{Bose}},
  \bibinfo{author}{\bibfnamefont{R.~E.} \bibnamefont{Angulo}},
  \bibinfo{author}{\bibfnamefont{L.}~\bibnamefont{Wang}},
  \bibinfo{author}{\bibfnamefont{W.~A.} \bibnamefont{Hellwing}},
  \bibinfo{author}{\bibfnamefont{J.~F.} \bibnamefont{Navarro}},
  \bibinfo{author}{\bibfnamefont{S.}~\bibnamefont{Cole}}, \bibnamefont{and}
  \bibinfo{author}{\bibfnamefont{C.~S.} \bibnamefont{Frenk}},
  \bibinfo{journal}{Mon. Not. Roy. Astron. Soc.}
  \textbf{\bibinfo{volume}{460}}, \bibinfo{pages}{1214} (\bibinfo{year}{2016}),
  \eprint{1601.02624}.

\bibitem[{\citenamefont{Simon}(2018)}]{Simon_2018}
\bibinfo{author}{\bibfnamefont{J.~D.} \bibnamefont{Simon}},
  \bibinfo{journal}{Astrophys. J.} \textbf{\bibinfo{volume}{863}},
  \bibinfo{pages}{89} (\bibinfo{year}{2018}).

\bibitem[{\citenamefont{Taylor and Navarro}(2001)}]{Taylor:2001bq}
\bibinfo{author}{\bibfnamefont{J.~E.} \bibnamefont{Taylor}} \bibnamefont{and}
  \bibinfo{author}{\bibfnamefont{J.~F.} \bibnamefont{Navarro}},
  \bibinfo{journal}{Astrophys. J.} \textbf{\bibinfo{volume}{563}},
  \bibinfo{pages}{483} (\bibinfo{year}{2001}), \eprint{astro-ph/0104002}.

\bibitem[{\citenamefont{Rocha et~al.}(2013)\citenamefont{Rocha, Peter, Bullock,
  Kaplinghat, Garrison-Kimmel, Onorbe, and Moustakas}}]{Rocha:2012jg}
\bibinfo{author}{\bibfnamefont{M.}~\bibnamefont{Rocha}},
  \bibinfo{author}{\bibfnamefont{A.~H.~G.} \bibnamefont{Peter}},
  \bibinfo{author}{\bibfnamefont{J.~S.} \bibnamefont{Bullock}},
  \bibinfo{author}{\bibfnamefont{M.}~\bibnamefont{Kaplinghat}},
  \bibinfo{author}{\bibfnamefont{S.}~\bibnamefont{Garrison-Kimmel}},
  \bibinfo{author}{\bibfnamefont{J.}~\bibnamefont{Onorbe}}, \bibnamefont{and}
  \bibinfo{author}{\bibfnamefont{L.~A.} \bibnamefont{Moustakas}},
  \bibinfo{journal}{Mon. Not. Roy. Astron. Soc.}
  \textbf{\bibinfo{volume}{430}}, \bibinfo{pages}{81} (\bibinfo{year}{2013}),
  \eprint{1208.3025}.

\bibitem[{\citenamefont{{Ascasibar} and
  {Gottl{\"o}ber}}(2008)}]{2008MNRAS.386.2022A}
\bibinfo{author}{\bibfnamefont{Y.}~\bibnamefont{{Ascasibar}}} \bibnamefont{and}
  \bibinfo{author}{\bibfnamefont{S.}~\bibnamefont{{Gottl{\"o}ber}}},
  \bibinfo{journal}{Mon.~Not.~Roy.~Astron.~Soc.}
  \textbf{\bibinfo{volume}{386}}, \bibinfo{pages}{2022} (\bibinfo{year}{2008}),
  \eprint{0802.4348}.

\bibitem[{\citenamefont{Colquhoun et~al.}(2021)\citenamefont{Colquhoun, Heeba,
  Kahlhoefer, Sagunski, and Tulin}}]{Colquhoun:2020adl}
\bibinfo{author}{\bibfnamefont{B.}~\bibnamefont{Colquhoun}},
  \bibinfo{author}{\bibfnamefont{S.}~\bibnamefont{Heeba}},
  \bibinfo{author}{\bibfnamefont{F.}~\bibnamefont{Kahlhoefer}},
  \bibinfo{author}{\bibfnamefont{L.}~\bibnamefont{Sagunski}}, \bibnamefont{and}
  \bibinfo{author}{\bibfnamefont{S.}~\bibnamefont{Tulin}},
  \bibinfo{journal}{Phys. Rev. D} \textbf{\bibinfo{volume}{103}},
  \bibinfo{pages}{035006} (\bibinfo{year}{2021}), \eprint{2011.04679}.

\end{thebibliography}

\end{document}